\pdfoutput=1
\documentclass[11pt,twoside,a4paper,cmspaper,final,collab]{cms-tdr}
\def\svnVersion{b4c26ae}\def\svnDate{2022/05/16}\def\cmsCernNoTag{CERN-EP-2021-252}\def\cmsCernDate{\today}\def\cmsMessage{Published in the Journal of High Energy Physics as \href{http://dx.doi.org/10.1007/JHEP06(2022)156}{\doi{10.1007/JHEP06(2022)156}.}}
\begin{document}\cmsNoteHeader{EXO-19-020}

\newcommand{\mt}{\ensuremath{m_{\mathrm{T}}}\xspace}
\newcommand{\MT}{\mt}
\newcommand{\RT}{\ensuremath{R_{\mathrm{T}}}\xspace}
\newcommand{\PZprime}{\ensuremath{{\PZ}^{\prime}}\xspace}
\newcommand{\mZprime}{\ensuremath{m_{\PZprime}}\xspace}
\newcommand{\sigmaZprime}{\ensuremath{\sigma_{\PZprime}}\xspace}
\newcommand{\wZprime}{\ensuremath{\Gamma_{\PZprime}}\xspace}
\newcommand{\Bdark}{\ensuremath{\mathcal{B}_{\text{dark}}}\xspace}
\newcommand{\mDark}{\ensuremath{m_{\text{dark}}}\xspace}
\newcommand{\aDark}{\ensuremath{\alpha_{\text{dark}}}\xspace}
\newcommand{\lamDark}{\ensuremath{\Lambda_{\text{dark}}}\xspace}
\newcommand{\aDarkPeak}{\ensuremath{\aDark^{\text{peak}}}\xspace}
\newcommand{\lamDarkPeak}{\ensuremath{\lamDark^{\text{peak}}}\xspace}
\newcommand{\aDarkHigh}{\ensuremath{\aDark^{\text{high}}}\xspace}
\newcommand{\aDarkLow}{\ensuremath{\aDark^{\text{low}}}\xspace}
\newcommand{\rinv}{\ensuremath{r_{\text{inv}}}\xspace}
\newcommand{\Pqdark}{\ensuremath{\chi}\xspace}
\newcommand{\mqdark}{\ensuremath{m_{\Pqdark}}\xspace}
\newcommand{\Paqdark}{\ensuremath{\overline{\chi}}\xspace}
\newcommand{\PqdarkO}{\ensuremath{\chi_{1}}\xspace}
\newcommand{\PqdarkT}{\ensuremath{\chi_{2}}\xspace}
\newcommand{\Ppidark}{\ensuremath{\pi_{\text{dark}}}\xspace}
\newcommand{\PpidarkDM}{\ensuremath{\Ppidark^{\mathrm{DM}}}\xspace}
\newcommand{\Prhodark}{\ensuremath{\rho_{\text{dark}}}\xspace}
\newcommand{\PrhodarkDM}{\ensuremath{\Prhodark^{\mathrm{DM}}}\xspace}
\newcommand{\ZprimeToDark}{\ensuremath{\PZprime \to \Pqdark\Paqdark}\xspace}
\newcommand{\gq}{\ensuremath{g_{\PQq}}\xspace}
\newcommand{\gqdark}{\ensuremath{g_{\Pqdark}}\xspace}
\newcommand{\gDM}{\ensuremath{g_{\text{DM}}}\xspace}
\newcommand{\Qdark}{\ensuremath{Q_\text{dark}}\xspace}
\newcommand{\Nc}{\ensuremath{N_{c}^\text{dark}}\xspace}
\newcommand{\Nf}{\ensuremath{N_{f}^\text{dark}}\xspace}
\newcommand{\mq}{\ensuremath{m_{\PQq}}\xspace}
\newcommand{\Nstable}{\ensuremath{N_{\text{stable}}}\xspace}
\newcommand{\Nunstable}{\ensuremath{N_{\text{unstable}}}\xspace}
\newcommand{\widejet}{\ensuremath{\mathrm{J}}\xspace}
\newcommand{\narrowjet}{\ensuremath{\mathrm{j}}\xspace}
\newcommand{\deta}{\ensuremath{\Delta\eta(\widejet_{1},\widejet_{2})}\xspace}
\newcommand{\mindphi}{\ensuremath{\Delta\phi_{\text{min}}}\xspace}
\newcommand{\zjets}{{{\PZ}+jets}\xspace}
\newcommand{\wjets}{{{\PW}+jets}\xspace}
\newcommand{\znnjets}{\ensuremath{{\PZ}{(\to\Pgn\Pagn)\text{+jets}}}\xspace}
\newcommand{\wlnjets}{\ensuremath{{\PW}{(\to\ell\Pgn)\text{+jets}}}\xspace}
\newcommand{\Imini}{\ensuremath{I_{\text{mini}}}\xspace}
\newcommand{\ptsub}[1]{\ensuremath{p_{\mathrm{T},#1}}\xspace}
\newcommand{\etsub}[1]{\ensuremath{E_{\mathrm{T},#1}}\xspace}
\newcommand{\ptvecsub}[1]{\ensuremath{\vec{p}_{\mathrm{T},#1}}\xspace}
\newcommand{\mjj}{\ensuremath{m_{\widejet\widejet}}\xspace}
\newcommand{\ptvecjj}{\ensuremath{\ptvecsub{\widejet\widejet}}\xspace}
\newcommand{\ptjj}{\ensuremath{\ptsub{\widejet\widejet}}\xspace}
\newcommand{\etjj}{\ensuremath{\etsub{\widejet\widejet}}\xspace}
\newcommand{\phijjmiss}{\ensuremath{\phi_{\widejet\widejet,\text{miss}}}\xspace}
\newcommand{\DRmax}{\ensuremath{\Delta R_{\text{max}}}\xspace}
\newcommand{\Pchhad}{\ensuremath{\mathrm{h}^{\pm}}\xspace}
\newcommand{\Pnhad}{\ensuremath{\mathrm{h}^{0}}\xspace}
\newcommand{\tauTO}{\ensuremath{\tau_{21}}\xspace}
\newcommand{\tauTT}{\ensuremath{\tau_{32}}\xspace}
\newcommand{\Ntwo}{\ensuremath{N_{2}^{(1)}}\xspace}
\newcommand{\Nthree}{\ensuremath{N_{3}^{(1)}}\xspace}
\newcommand{\msd}{\ensuremath{m_{\mathrm{SD}}}\xspace}
\newcommand{\axmajor}{\ensuremath{\sigma_{\text{major}}}\xspace}
\newcommand{\axminor}{\ensuremath{\sigma_{\text{minor}}}\xspace}
\newcommand{\ptD}{\ensuremath{D_{\pt}}\xspace}
\newcommand{\abseta}{\ensuremath{\abs{\eta}}\xspace}
\newcommand{\cdead}{\ensuremath{c_{\text{nonfunctional}}}\xspace}
\newcommand{\DRphispike}{\ensuremath{\Delta R(\narrowjet_{1,2},\cdead)}\xspace}
\newcommand{\nmuons}{\ensuremath{N_{\Pgm}}\xspace}
\newcommand{\neles}{\ensuremath{N_{\Pe}}\xspace}
\newcommand{\dphij}{\ensuremath{\Delta\phi(\vec{\widejet},\ptvecmiss)}\xspace}
\newcommand{\girth}{\ensuremath{g_{\text{jet}}}\xspace}
\newcommand{\bkgrej}{\ensuremath{1/\varepsilon_{\text{bkg}}}\xspace}
\newcommand{\sigeff}{\ensuremath{\varepsilon_{\text{sig}}}\xspace}
\newcommand{\rinj}{\ensuremath{\sigma_\text{injected}}\xspace}
\newcommand{\rext}{\ensuremath{\sigma_\text{extracted}}\xspace}
\newcommand{\rerr}{\ensuremath{\varepsilon_{\rext}}\xspace}
\newcommand{\sqrts}{\ensuremath{\sqrt{s}}\xspace}
\newcommand{\sigmaexp}{\ensuremath{\sigma_{\text{exp}}}\xspace}
\newcommand{\colspace}{\hphantom{\ensuremath{=0.3)}}}
\newlength\cmsTabSkip\setlength{\cmsTabSkip}{1ex}

\newcolumntype{R}{>{$}r<{$}}
\newcolumntype{L}{>{$}l<{$}}
\newcolumntype{M}{R@{$\;$}L}

\cmsNoteHeader{EXO-19-020}
\title{Search for resonant production of strongly coupled dark matter in proton-proton collisions at 13\TeV}

\date{\today}

\abstract{
The first collider search for dark matter arising from a strongly coupled hidden sector is presented and uses a data sample corresponding to 138\fbinv, collected with the CMS detector at the CERN LHC, at $\sqrt{s}=13\TeV$. The hidden sector is hypothesized to couple to the standard model (SM) via a heavy leptophobic \PZprime mediator produced as a resonance in proton-proton collisions. The mediator decay results in two ``semivisible'' jets, containing both visible matter and invisible dark matter. The final state therefore includes moderate missing energy aligned with one of the jets, a signature ignored by most dark matter searches. No structure in the dijet transverse mass spectra compatible with the signal is observed. Assuming the \PZprime boson has a universal coupling of 0.25 to the SM quarks, an inclusive search, relevant to any model that exhibits this kinematic behavior, excludes mediator masses of 1.5--4.0\TeV at 95\% confidence level, depending on the other signal model parameters. To enhance the sensitivity of the search for this particular class of hidden sector models, a boosted decision tree (BDT) is trained using jet substructure variables to distinguish between semivisible jets and SM jets from background processes. When the BDT is employed to identify each jet in the dijet system as semivisible, the mediator mass exclusion increases to 5.1\TeV, for wider ranges of the other signal model parameters. These limits exclude a wide range of strongly coupled hidden sector models for the first time.
}

\hypersetup{%
pdfauthor={CMS Collaboration},%
pdftitle={Search for resonant production of strongly coupled dark matter in proton-proton collisions at 13 TeV},%
pdfsubject={CMS},%
pdfkeywords={CMS, dark matter, beyond standard model}}

\maketitle

\section{Introduction}\label{sec:intro}

{\tolerance=800
Over the past several decades, precision collider studies of elementary particles, such as the electroweak bosons, the top quark, and the recently discovered Higgs boson,
have demonstrated remarkable agreement with theoretical calculations from the standard model (SM) of particle physics.
However, many astronomical observations---including galaxy rotation curves~\cite{Rubin:1980zd,Persic:1995ru},
strong (weak) gravitational lensing observations of galaxy cluster collisions~\cite{Clowe:2006eq} (large-scale structures~\cite{Chang:2017kmv}),
and the cosmic microwave background power spectrum~\cite{Planck:2018vyg}---indicate the existence and prevalence of dark matter (DM).
The SM is not sufficient to explain these observations, which provides a compelling motivation to search for DM in the context of physics beyond the SM.
\par}

Various extensions to the SM that include DM and are compatible with the astronomical data
allow proton-DM scattering, DM-DM annihilation, or DM production at colliders.
The possible masses of the DM particles and the particles that mediate the proton-DM interactions span at least 36 orders of magnitude~\cite{Battaglieri:2017aum}.
Collider searches have primarily focused on models with weakly interacting massive particles (WIMPs).
These WIMPs are typically predicted to fall in a favorable mass range for collider production~\cite{Jungman:1995df}
and present relatively clear signatures with missing momentum recoiling against visible SM particles.
The CMS experiment has examined such signatures using different visible objects,
including a jet or vector boson~\cite{CMS:2018ffd,CMS:2020ulv,CMS:2021far}, a Higgs boson~\cite{Sirunyan:2018gdw}, a top quark~\cite{Sirunyan:2019gfm},
or jets from vector boson fusion in the case of a Higgs portal~\cite{Sirunyan:2018owy}.
These publications, along with similar searches from the ATLAS experiment~\cite{Aaboud:2018xdl,Aaboud:2018sfi,ATLAS:2019wdu,ATLAS:2021hza,ATLAS:2020xzu,ATLAS:2021kxv}, have reported no evidence for WIMP dark matter.

Hidden valley (HV) theories~\cite{Strassler:2006im} are alternative scenarios that propose dark sectors with potentially multiple new particles and new forces, decoupled from the SM except for mediator particles.
Searches conducted by the LHCb, ATLAS, and CMS experiments have not found evidence for models with dark photons, predicted by the simplest dark sectors with a new $U(1)$
force~\cite{Aaij:2017rft,Aaboud:2018jbr,Sirunyan:2018mgs,Sirunyan:2019xst,Aad:2019tua,Aaij:2019bvg,Sirunyan:2019wqq,Aaij:2020ikh,CMS:2020krr}.
Alternatively, the dark sector may contain a new confining force $SU(N)$, an analog to quantum chromodynamics (QCD) in the SM.
These models are largely unexplored in experimental searches; they generate dark showers, which present a wide range of novel kinematic signatures~\cite{Alimena:2019zri}.

Several considerations motivate the search for a strongly coupled dark sector at the LHC.
The abundance of visible matter arises from a baryon asymmetry and a similar mechanism may explain the abundance of dark matter~\cite{Petraki:2013wwa}.
Cosmological measurements of the dark matter density have established it to be of the same order as that of the visible matter, roughly a factor of five larger~\cite{Ade:2015xua}.
This observed similarity with visible matter suggests that dark matter may consist of composite particles.
In some cases, the scale of the new confining force, also called dark QCD, may be related to SM QCD~\cite{Bai:2013xga},
favoring scales on the order of 10\GeV and mediator masses at the \TeVns scale.
Generically, these models can produce the correct DM relic density~\cite{Beauchesne:2018myj}.

\begin{figure}[htb!]
\centering
\includegraphics[width=0.49\linewidth]{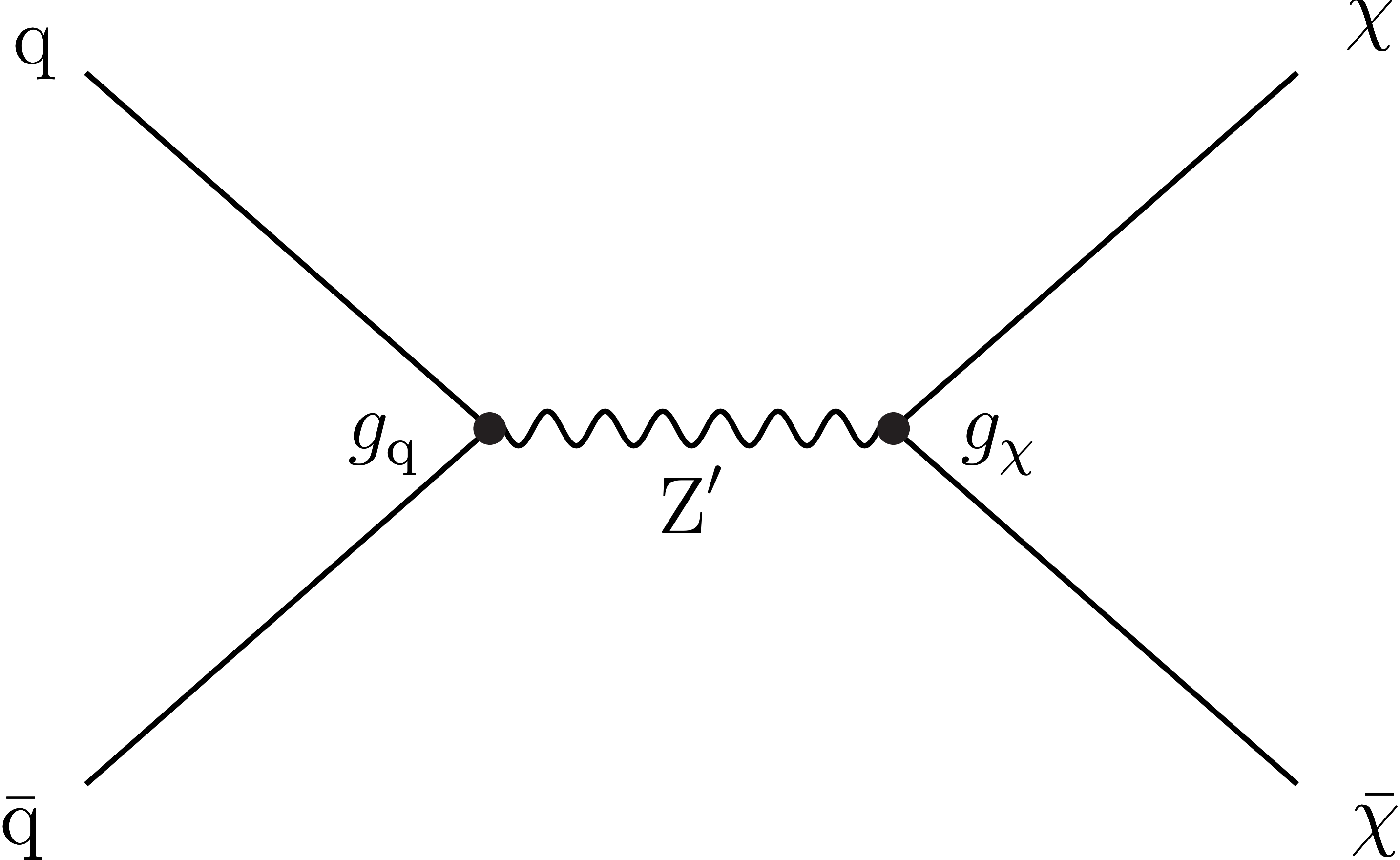}
\caption{Feynman diagram of leading-order resonant production of dark quarks through a \PZprime mediator.
The relevant couplings to SM quarks and dark quarks, \gq and \gqdark, are indicated at the labeled vertices.
}
\label{fig:s-chan-diagram}
\end{figure}

In this paper, we consider the class of models proposed in Refs.~\cite{Cohen:2015toa,Cohen:2017pzm},
in particular the resonant production process $\PQq\PAQq \to \ZprimeToDark$ depicted in Fig.~\ref{fig:s-chan-diagram}.
This process involves a leptophobic \PZprime boson mediator arising from a broken $U(1)$ symmetry, with couplings to the SM quarks \gq and dark quarks \gqdark.
The dark sector contains several flavors of dark quarks (\PqdarkO, \PqdarkT, \ldots) that form bound states called dark hadrons, which may be either stable or unstable.
The unstable dark hadrons decay promptly to SM quarks, while the stable dark hadrons are DM candidates that traverse the detector without interacting.
This leads to collimated mixtures of visible and invisible particles, known as ``semivisible'' jets (SVJ).
The heavy \PZprime boson tends to be produced at rest, resulting in two jets that are back-to-back in the transverse plane.
The total amount of missing transverse momentum is expected to be moderate because both jets contain invisible particles,
so a portion of the transverse component of the overall missing momentum will cancel.
Therefore, the signature of this model is a pair of jets along with the missing transverse momentum that is aligned with one of the jets.
The jets are expected to be wider than typical SM jets because they arise from a multi-step process:
the dark quarks shower and hadronize in the dark sector, the unstable dark hadrons decay to SM quarks, and finally the SM quarks shower and hadronize visibly.
The SM quarks are much less massive than the dark hadrons, so they are produced in the intermediate decay step with significantly higher relative momentum, spreading out the SM particle showers.

The fraction of stable, invisible dark hadrons, $\rinv = \Nstable/(\Nstable+\Nunstable)$, can take any value between 0 and 1, as described in Ref.~\cite{Cohen:2015toa}.
Since events containing jets aligned with missing transverse momentum are explicitly rejected from the signal regions of existing collider DM searches,
a significant portion of the parameter space of this model is not covered, particularly intermediate values of \rinv.
The current DM searches are primarily sensitive to dark sector models with $\rinv \approx 1$ that yield events with high missing momentum along with some initial-state radiation.
The current dijet searches~\cite{Sirunyan:2019vgj,Aad:2019hjw} are sensitive to dark sector models with $\rinv \approx 0$ that yield events with two jets and low missing momentum.
Therefore, the search presented here complements these efforts.
The case where the unstable dark hadrons decay after some finite lifetime results instead in emerging jets~\cite{Schwaller:2015gea},
for which a search has already been conducted by CMS~\cite{Sirunyan:2018njd} using 13\TeV data corresponding to an integrated luminosity of 16.1\fbinv.
That search excluded mediator particles with masses up to 1.25\TeV for dark hadrons with masses between 1 and 10\GeV and decay lengths between 5 and 225\mm.

We analyze data from proton-proton ($\Pp\Pp$) collisions delivered by the CERN LHC in 2016--2018 and collected with the CMS detector.
The data sample corresponds to an integrated luminosity of 138\fbinv~\cite{CMS:2021xjt}.
The search is conducted following a dual strategy, in which we present two sets of results.
First, a generic or inclusive search is sensitive to any model that presents the signature of jets aligned with missing transverse momentum.
Second, a dedicated search for the specific class of dark sector models introduced above
uses a boosted decision tree (BDT) to distinguish semivisible jets from SM jets using jet substructure variables.
The SM background is dominated by events in which a mismeasured jet causes artificial missing momentum.
These events are composed only of jets produced through the strong interaction, which we will describe as the QCD multijet process.
There are additional background contributions from the \ttbar, \wlnjets, and \znnjets processes, which produce events with genuine missing momentum associated with neutrinos.
Tabulated results are provided in the HEPData record for this analysis~\cite{hepdata}.

\section{The CMS detector}\label{sec:cms}

The central feature of the CMS apparatus is a superconducting solenoid of 6\unit{m} internal diameter, providing a magnetic field of 3.8\unit{T}.
Within the solenoid volume are a silicon pixel and strip tracker, a lead tungstate crystal electromagnetic calorimeter (ECAL),
and a brass and scintillator hadron calorimeter (HCAL), each composed of a barrel and two endcap sections.
The ECAL comprises 75,848 readout channels, of which 1.0--1.4\% were nonfunctional during the LHC data-taking period considered in this paper.
Forward calorimeters extend the pseudorapidity coverage provided by the barrel and endcap detectors.
Muons are measured in gas-ionization detectors embedded in the steel flux-return yoke outside the solenoid.
Events of interest are selected using a two-tiered trigger system.
The first level, composed of custom hardware processors,
uses information from the calorimeters and muon detectors to select events at a rate of around 100\unit{kHz} within a fixed latency of about 4\mus~\cite{CMS:2020cmk}.
The second level, known as the high-level trigger, consists of a farm of processors running a version of the full event reconstruction software optimized for fast processing,
and reduces the event rate to around 1\unit{kHz} before data storage~\cite{CMS:2016ngn}. 
A more detailed description of the CMS detector, together with a definition of the coordinate system used and the relevant kinematic variables, can be found in Ref.~\cite{Chatrchyan:2008zzk}. 

\section{Signal model}\label{sec:sig}

The implementation of the \PZprime signal HV model is based on Ref.~\cite{Cohen:2015toa}.
The mediator mass \mZprime and couplings \gq and \gqdark are undetermined parameters.
The production cross section \sigmaZprime is determined by \mZprime and \gq,
while the branching fraction $\Bdark = \mathcal{B}(\ZprimeToDark)$ depends on both \gq and \gqdark.
The effective cross section is defined as the product $\sigmaZprime\Bdark$.
The dark QCD force is carried by a dark gluon, and binds the dark quarks into dark hadrons, which can be pseudoscalars ($\pi$) or vectors ($\rho$).
The unstable dark hadrons are \Ppidark and \Prhodark, respectively, while the stable dark hadrons are \PpidarkDM and \PrhodarkDM.
The dark hadrons are assumed to be degenerate in mass, with the mass scale \mDark as another parameter.
The coupling strength of the dark QCD force, \aDark, is also undetermined; its value primarily influences the dark shower dynamics
by modifying the multiplicity and the transverse momentum \pt of the dark hadrons, as well as the width of the resulting spray of particles.
The last parameter of interest is the fraction of invisible dark hadrons, \rinv.
The variation of this fraction is not considered in other theories, making it the most novel parameter.
In summary, the leptophobic \PZprime model has five parameters: the effective cross section, \mZprime, \mDark, \aDark, and \rinv.

For the search presented here, we restrict the range of the mass scale \mDark between 1 and 100\GeV; the lower bound is determined by modeling constraints in the \PYTHIA~\cite{Sjostrand:2014zea} generator,
while the upper bound ensures enough hadrons are produced to form a dark shower and the corresponding semivisible jet.
The mass of the dark quark species is fixed at ${\mqdark = \mDark/2}$, following Ref.~\cite{Cohen:2015toa}.
The running coupling \aDark can also be expressed in terms of the dark coupling scale \lamDark, which has dimensions of energy, as
$\aDark(\lamDark) = \pi/\left(b_{0}\ln\left(\Qdark/\lamDark\right)\right)$,
with the factor $b_{0} = (11\Nc - 2\Nf)/6$.
Here, we choose the number of dark colors $\Nc = 2$, the number of dark flavors $\Nf = 2$, and the scale $\Qdark = 1\TeV$.
The effects from changing \lamDark depend strongly on the dark hadron mass,
so we choose the scale to maximize the number of dark hadrons in the shower for each \mDark value.
An analytic solution for this requirement is not known,
so an approximate numerical relationship $\lamDarkPeak = 3.2(\mDark)^{0.8}\GeV$, with \mDark in {\GeVns}, is obtained using \PYTHIA.
The benchmark value \aDarkPeak is computed from \lamDarkPeak with the above equation, and we consider variations of ${\pm}50\%$ (\aDarkLow, \aDarkHigh).
The smallest values for \Nc and \Nf that reproduce the expected behavior of the model---the formation of both stable and unstable dark hadrons---are chosen.
Hence, the dark QCD is a confining $SU(2)$ force in this model.
We choose ${\gq = 0.25}$ and ${\gqdark = 1.0/\sqrt{\smash[b]{\Nc\Nf}} = 0.5}$ as the values for the \PZprime boson couplings.
With these choices, the production cross section, branching fraction, and mediator width are compatible with the benchmark model recommended by the LHC DM Working Group~\cite{Albert:2017onk}.
In particular, the branching fraction is ${\Bdark = 47\%}$ and the mediator width is $\wZprime/\mZprime = 5.6\%$,
for which the narrow width approximation holds~\cite{Boveia:2016mrp}.
The difference between the \gqdark value and the LHC DM coupling benchmark value ${\gDM = 1.0}$ accounts for the multiple flavors and colors of dark quarks.

Several features of the HV model are implemented by modifying \PYTHIA parameters for particle branching fractions.
The dark hadrons may be stable or unstable.
The unstable vector dark hadrons \Prhodark can decay to a pair of SM quarks of any flavor with equal probability, as long as $\mDark \geq 2\mq$.
In contrast, the unstable pseudoscalar dark hadrons \Ppidark must decay through a mass insertion, and therefore decay preferentially to more massive SM quarks~\cite{Cohen:2015toa}.
This preference is reflected in their branching fractions, which are calculated including the effect of quark mass running~\cite{Ellis:1996mzs}.
To simulate the stable dark hadrons, we reduce the dark hadron branching fraction to SM quarks according to the value of the \rinv parameter.
The probability of producing a vector dark hadron is set to 0.75.
Finally, a $\mathbb{Z}_{2}$ symmetry is explicitly enforced by rejecting events with an odd number of stable dark hadrons.

\section{Simulation}\label{sec:samples}

We use simulated samples to model signal and SM background processes.
Because of changes in detector conditions, including detector upgrades and aging, we use separate samples
for each year of data taking: 2016, 2017, and 2018.
The detector response to simulated particles is modeled using the \GEANTfour software~\cite{Agostinelli:2002hh}.
Custom simulations of the detector electronics are used to produce readouts similar to those observed in data.
Multiple $\Pp\Pp$ interactions (pileup) are also included in the simulation.
The simulated samples are corrected to make the pileup distribution match the distribution in data as closely as possible.

The \MGvATNLO event generator~\cite{Alwall:2014hca} with the MLM matching procedure~\cite{Alwall:2007fs} is used to simulate the \ttbar, $\wlnjets$, and $\znnjets$ processes
with leading-order (LO) matrix element calculations including up to three, four, and four additional partons, respectively.
The 2016 samples were generated with \MGvATNLO\,2.2.2 and the 2017 and 2018 samples with \MGvATNLO\,2.4.2.
These samples are normalized to the next-to-next-to-leading-order (NNLO)
cross sections~\cite{Beneke:2011mq,Cacciari:2011hy,Baernreuther:2012ws,Czakon:2012zr,Czakon:2012pz,Czakon:2013goa,Li:2012wna}.
The \PYTHIA generator is used to simulate the QCD multijets process at LO as a $2{\to}2$ interaction,
with versions 8.212, 8.226, and 8.230 employed for the 2016, 2017, and 2018 simulated samples, respectively.
The QCD multijet simulation is normalized to the LO cross section provided by \PYTHIA,
and an empirically estimated correction of 1.2--1.9, taking into account different detector conditions during the three data-taking periods, is applied
to account for the difference in the order of the cross section computation relative to the other backgrounds.
The same versions of \PYTHIA are used to simulate parton showering and hadronization for all SM processes.
For the samples associated with the 2016 data set, those generated with \MADGRAPH use the NNPDF3.0 LO~\cite{NNPDF:2014otw} parton distribution functions (PDF)
and those generated with \PYTHIA use NNPDF2.3 LO~\cite{Ball:2013hta}; all of them use the CUETP8M1~\cite{Khachatryan:2015pea} underlying-event tune for \PYTHIA.
The 2017 and 2018 samples use the NNPDF3.1 NNLO PDFs~\cite{Ball:2017nwa} and the CP5 tune~\cite{Sirunyan:2019dfx}.
The simulations of SM background processes are used to compare to signal in order to optimize various kinematic criteria for sensitivity,
to check the agreement with data for basic kinematic variables, and to train the BDT.

The simulated signal samples are also generated with \PYTHIA, using version 8.226 for 2016 and 8.230 for 2017 and 2018.
The dedicated HV module is used for the showering and hadronization of the dark hadrons.
Compared to the SM background simulation, newer versions of \PYTHIA are used for the signal simulation in order to incorporate essential features, such as the running of the dark coupling.
The 2016 samples use the NNPDF2.3 LO PDF and the CUETP8M1 tune, and
the 2017 and 2018 samples use the NNPDF3.1 LO PDF~\cite{Ball:2017nwa} and the CP2 tune~\cite{Sirunyan:2019dfx}.
The samples are normalized using next-to-leading-order cross sections computed with a universal coupling between the \PZprime boson and the SM quarks.

As described in Section~\ref{sec:sig}, the signal model has five free parameters,
four of which affect the physical properties of the resulting events,
while the effective cross section just affects the total yield of events containing semivisible jets.
We define the benchmark values for these parameters as $\mZprime = 3.1\TeV$,
$\mDark = 20\GeV$, $\rinv = 0.3$, and $\aDark = \aDarkPeak$, where $\aDarkPeak \approx 0.313$ for $\mDark = 20\GeV$.
These values are chosen to have high acceptance for the selection defined in Section~\ref{sec:evtsel}.
To keep the number of concrete models to a reasonable level,
we generate simulated samples varying only two parameters at a time,
producing three two-dimensional ``scans'' of the model parameter space shown in Table~\ref{tab:sigscan}.
These scans will be used to interpret the results of the search,
with the search variable distributions for each signal model formed directly from the simulated events.

\begin{table}[htb]
\topcaption{The three two-dimensional signal model parameter scans. For each scan, the ranges of the parameters that are varied in that scan are indicated by dashes.}
\centering
\renewcommand{\arraystretch}{1.3}
\begin{tabular}{lcccc}
Scan & \mZprime [\TeVns{}] & \mDark [\GeVns{}] & \rinv & \aDark \\
\hline
1 & 1.5--5.1 & 1--100 & 0.3 & \aDarkPeak \\
2 & 1.5--5.1 & 20 & 0--1 & \aDarkPeak \\
3 & 1.5--5.1 & 20 & 0.3 & \aDarkLow--\aDarkHigh \\
\end{tabular}
\label{tab:sigscan}
\end{table}

\section{Event reconstruction and triggering}\label{sec:reco}

The events recorded with the CMS detector are reconstructed using the particle-flow (PF) algorithm~\cite{CMS-PRF-14-001}, which aims to identify every particle in each event.
The PF algorithm combines information from all subdetector systems in an optimized manner.
Each reconstructed particle, also called a PF candidate, is identified as a charged hadron, electron, muon, neutral hadron, or photon.
The anti-\kt algorithm~\cite{Cacciari:2008gp, Cacciari:2011ma} is used to cluster the PF candidates into jets.
The missing transverse momentum vector \ptvecmiss is computed as the negative vector \ptvec sum of the PF candidates in an event;
its magnitude is denoted as \ptmiss~\cite{Sirunyan:2019kia}.
In order to improve the quality of the PF reconstruction, additional identification criteria are applied to electron~\cite{CMS:2020uim} and muon candidates~\cite{Sirunyan:2018fpa},
which are required to have $\pt > 10\GeV$ and $\abseta < 2.4$.

The candidate vertex with the largest value of summed physics-object $\pt^2$ is taken to be the primary $\Pp\Pp$ interaction vertex,
where the physics objects used for this determination are the jets and the missing transverse momentum.
For the vertex calculation, the tracks assigned to each vertex serve as input to the jet clustering algorithm,
and the missing transverse momentum is the negative vector \ptvec sum of the resulting jets.
Only charged-particle tracks associated with the primary vertex are considered when reconstructing the final collection of PF candidates.
Rejecting tracks associated with other vertices reduces the impact of pileup.

Two types of jets are used in this search. The primary collection of jets with $\pt>200\GeV$ is reconstructed using a distance parameter of $R = 0.8$ for the clustering algorithm,
because the signal jets are expected to have a broader radiation pattern than the SM background jets.
We denote these reconstructed objects with a capital \widejet, and we refer to them simply as ``jets'' in the rest of the paper.
The pileup-per-particle identification (PUPPI) algorithm~\cite{Bertolini:2014bba} is used to mitigate the effect of pileup on the jets.
The PUPPI algorithm makes use of a local shape variable that reflects the collinear versus soft diffuse structure in the neighborhood of the particle.
With this variable, as well as with event pileup properties and tracking information at the reconstructed particle level,
the algorithm computes the probability that a given neutral PF candidate results from pileup. The candidate momentum is multiplied by this probability~\cite{CMS:2020ebo}.
The jets are further corrected to account for nonlinearities in the detector energy response~\cite{Khachatryan:2016kdb}.
In simulated samples, the jet \pt is smeared using measured resolutions to match the observed data.
A set of quality criteria is applied to reject spurious jets arising from instrumental sources~\cite{CMS-PAS-JME-16-003}.

The jet clustering algorithm is also employed with a distance parameter of $R = 0.4$
to produce a second collection of jets, which are then required to have $\pt>30\GeV$.
We denote these reconstructed objects with a lowercase \narrowjet, and we refer to them as ``narrow jets'' in the rest of the paper.
The narrow jets are employed in the trigger and are also used to mitigate instrumental backgrounds, which are unrelated to the expected size and scale of the signal jets.
Energy corrections are also applied to the narrow jets, with an area-based pileup subtraction~\cite{Cacciari:2007fd,CMS:2020ebo} used in place of the PUPPI algorithm.
The effect of the narrow-jet energy corrections is propagated to the missing transverse momentum.

The data set considered in this paper is collected using triggers that place requirements on the narrow-jet \pt or on the \HT,
which is the scalar \pt sum of all narrow jets with $\pt>30\GeV$ and $\abseta < 3.0$.
The triggers used in 2016 (2017--2018) require a jet with $\pt > 450\,(500)\GeV$ or $\HT > 900\,(1050)\GeV$.
The thresholds were raised in 2017 to compensate for higher instantaneous luminosity.
The efficiency for an event to pass any of these trigger conditions is measured in data
using a data set collected with an independent trigger that requires a muon with $\pt > 50\GeV$.
The selection requirements that ensure a high trigger efficiency are described in the next section.

\section{Event selection}\label{sec:evtsel}

We select events with at least two jets.
Any event in which the two highest \pt jets are not both within $\abseta < 2.4$ is rejected.
Events with missing transverse momentum that have substantial energy in the endcap region are significantly more likely to originate from beam halo or noise induced by the radiation damage,
rather than from signal events, which tend to populate the barrel region.
Therefore, this requirement eliminates several sources of instrumental background, while having a negligible impact on the signal efficiency.

The dijet transverse mass \mt is used as the search variable.
In signal processes, it forms a peak with a kinematic endpoint at \mZprime and therefore approximately reconstructs the mass of the \PZprime mediator,
while in background processes, it is expected to have a smoothly falling spectrum.
This can be observed in Figs.~\ref{fig:prefit_cut}--\ref{fig:prefit_bdt} in Section~\ref{sec:bkg}.
The dijet transverse mass is computed from the four-vector of the dijet system and the \ptvecmiss~\cite{Cohen:2015toa}:
\begin{equation}\label{eq:mt}
        \begin{aligned}
\mt^{2} &= \left[\etjj + \ETmiss\right]^{2} - \left[\ptvecjj + \ptvecmiss\right]^{2} \\
        &= \mjj^{2}+2\ptmiss\left[\sqrt{\mjj^{2}+\ptjj^2} - \ptjj\cos(\phijjmiss)\right].
\end{aligned}
\end{equation}
Here, \mjj is the invariant mass of the system composed of the two highest \pt jets, and \ptvecjj is the vector sum of their \ptvec.
The quantity $\etjj^2$ is defined as $\mjj^2 + \abs{\ptvecjj}^2$, while we assume that there is just one massless, invisible particle in the final state, implying the missing transverse energy ${\ETmiss = \ptmiss}$.
This enables the simplification in the second line of Eq.~\eqref{eq:mt}, with \phijjmiss as the azimuthal angle between the dijet system and the missing transverse momentum.
Bins of width 100\GeV are used when measuring the \mt distribution; this width provides sufficiently populated bins for the majority of the spectrum.

\begin{figure*}[htb!]
\centering
\includegraphics[width=0.49\linewidth]{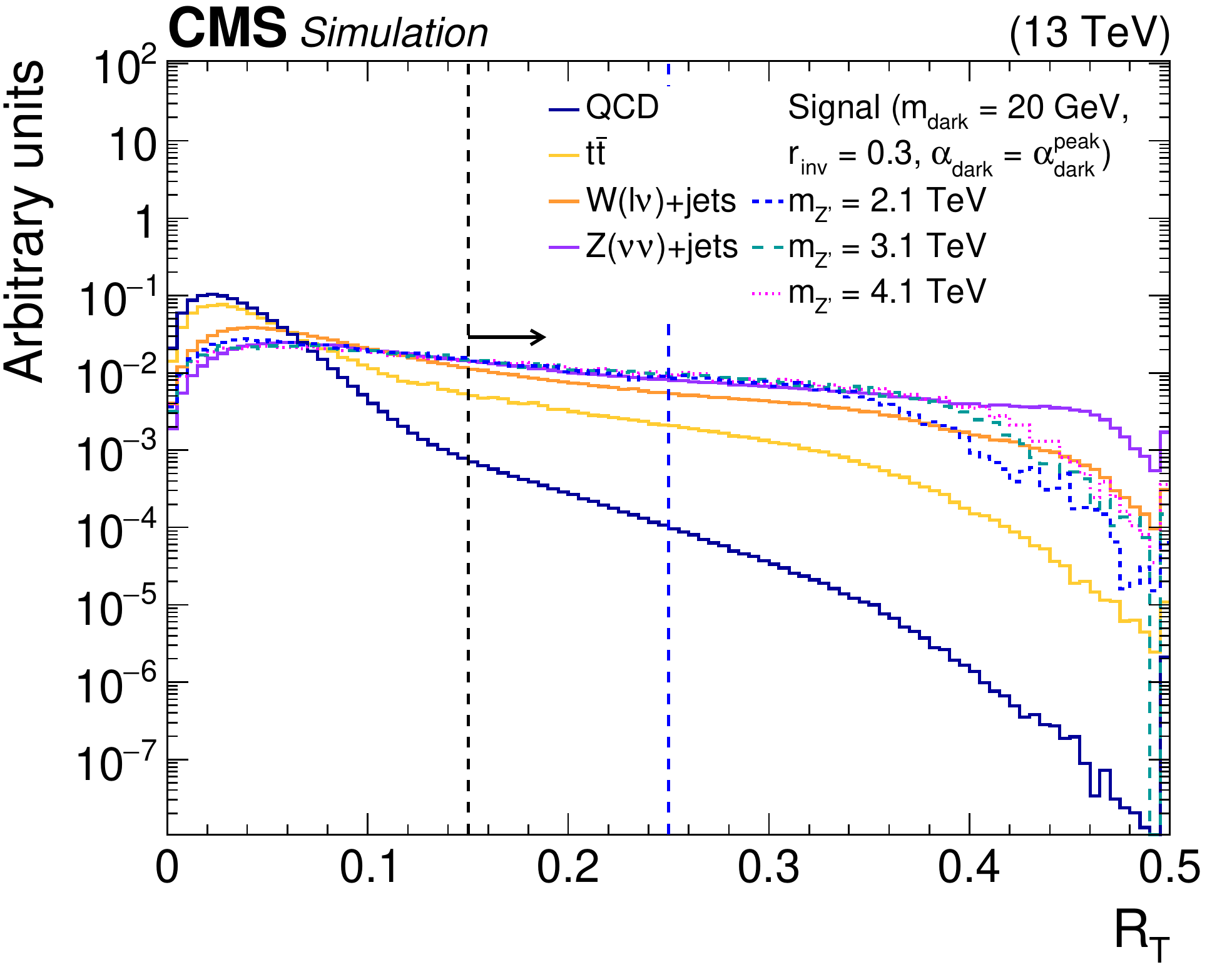}
\includegraphics[width=0.49\linewidth]{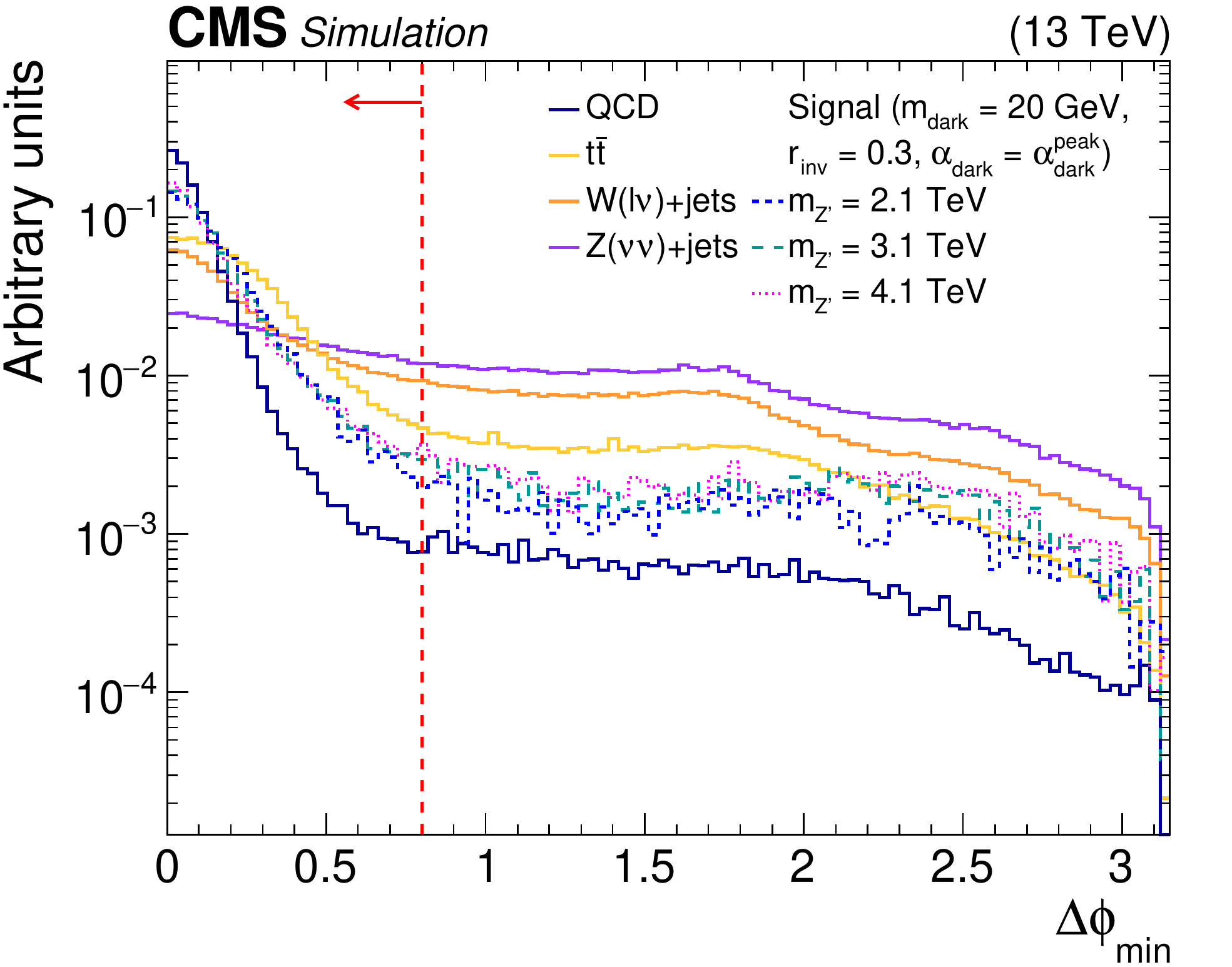}
\caption{The normalized distributions of the characteristic variables \RT and \mindphi
for the simulated SM backgrounds and several signal models.
For each variable, the requirement on that variable is omitted, but all other preselection requirements are applied.
The black (red) vertical dotted line indicates the preselection (final selection) requirement on the variable, if any.
The blue vertical dotted line indicates the boundary between different signal regions.
The last bin of each histogram includes the overflow events.
}
\label{fig:nMinus1}
\end{figure*}

The transverse ratio is defined as $\RT = \ptmiss/\mt$ and used in the selection instead of \ptmiss to identify events with invisible particles.
This is because \ptmiss is correlated with \mt, and therefore placing a requirement directly on \ptmiss would shift the peak of the background distribution to higher \mt values,
as well as causing the distribution to have a shallower slope.
The requirement $\RT > 0.15$, shown in Fig.~\ref{fig:nMinus1} (left), rejects 99\% of simulated QCD background events.
In particular, this requirement removes the majority of $t$-channel QCD events, which have higher \mt than $s$-channel events,
but still low \ptmiss, and therefore lower \RT values.
Accordingly, the proportion of background events from the \ttbar, \wjets, and \zjets processes, collectively described as the electroweak background, is enhanced.

The remaining $t$-channel QCD events have low trigger efficiency, because their high \mt values arise
from the large $\deta = \abs{\eta_{\widejet_{1}}-\eta_{\widejet_{2}}}$ separation between the two highest \pt jets, rather than from the jet \pt that is the basis of the trigger.
To reject these events, we additionally require $\deta < 1.5$.
With these requirements applied, the efficiency of the jet-based triggers is measured in the data to have a plateau at 98\% for $\MT > 2.0\TeV$.
Events with $\MT > 1.5\TeV$ are selected to be in the fully efficient region, defined as an efficiency of at least 95\% of the plateau value,
in order to ensure that the background events will have a falling spectrum.
The simulated signal and background events are corrected to match the measured trigger efficiency for each year of data taking.

In order to reduce the \ttbar and \wlnjets backgrounds, we veto events containing electrons or muons passing the kinematic and identification criteria noted in Section~\ref{sec:reco},
along with an additional requirement that the lepton candidates are isolated from other particles.
In other words, we require the number of electrons \neles to equal 0 and the number of muons \nmuons to equal 0.
Isolation is applied to reduce the contributions from jets misidentified as leptons, or leptons arising from hadron decays rather than prompt {\PW} boson decays.
It is quantified using the relative variable
$\Imini(\ell) = (\sum_{\DRmax}{\ptsub{\Pchhad}}+\text{max}[0,\sum_{\DRmax}{\ptsub{\gamma}+\ptsub{\Pnhad}-\ptsub{\text{pileup}}/2}])/\ptsub{\ell}$,
where the different particle candidates are denoted by $\ell$ for the charged lepton, $\Pchhad$ for charged hadrons, $\gamma$ for photons, and $\Pnhad$ for neutral hadrons.
The label ``pileup'' in the last term denotes charged hadrons that do not originate from the primary vertex and are used to estimate additional energy coming from neutral candidates whose vertex cannot be determined.
All the sums consider candidates within a cone of $\Delta R = \sqrt{\smash[b]{(\Delta\eta)^2+(\Delta\phi)^2}} < \DRmax$,
where \DRmax is 0.2 for $\ptsub{\ell} \leq 50\GeV$, $10\GeV/\ptsub{\ell}$ for $50 < \ptsub{\ell} < 200\GeV$, and 0.05 for $\ptsub{\ell} \geq 200\GeV$.
The radius of the cone decreases with increasing charged-lepton \pt in order to account for the effect of the Lorentz boost of the lepton's parent particle;
a larger boost leads to more collimated decay products~\cite{Rehermann:2010vq}.
We require $\Imini(\Pgm) < 0.4$ or $\Imini(\Pe) < 0.1$.
The lepton vetoes reject 40\% of the \ttbar and \wlnjets background events that remain after applying the selection requirements described previously in this section.
The events that cannot be rejected contain ``lost'' leptons that fail some aspect of the kinematic, identification, or isolation criteria.

Events with anomalously high \ptmiss values can occur due to a variety of reconstruction failures, detector malfunctions, or other noncollision backgrounds.
These events are rejected by dedicated filters designed to identify at least 85\% of the anomalous events, while having a false positive rate less than 0.1\%~\cite{Sirunyan:2019kia}.
The leading source of instrumental background is mismeasurement of jet energy because of nonfunctional ECAL readout channels (\cdead).
We employ a filter, optimized for the kinematic phase space of this search, to identify the nonfunctional channels.
This approach relies on the fact that the production of jets in physical SM multijet processes is expected to be isotropic in $\phi$.
Because of the \RT requirement for nonnegligible missing transverse momentum,
multijet background events will be preferentially selected if one of the jets is mismeasured because it overlaps with a nonfunctional channel.
Therefore, regions with nonfunctional channels can be identified as enhancements in the $\eta$-$\phi$ distributions of the two highest \pt narrow jets in each event.
In this custom filter, an enhancement is defined as a bin having a value approximately five standard deviations larger than the mean value for the distribution.
Events are rejected if either of the two highest \pt narrow jets overlaps with a nonfunctional channel identified in this way, determined by $\DRphispike < 0.1$.
This custom filter eliminates 40\% of the remaining QCD background,
while only reducing the signal efficiency by approximately 5\%.

The requirements described above constitute the initial selection or ``preselection''.
Preselected events are used to verify basic agreement in kinematic variables between data and simulation,
as well as to train the semivisible jet identification algorithm, described in Section~\ref{sec:tagger}.
As the final selection, we apply the three additional criteria described below that further enhance the signal relative to the backgrounds.

First, an unphysical contribution from misreconstructed jets, which are found to produce events with artificially high \MT values,
is rejected by vetoing events where the leading narrow jet has a photon energy fraction $f_{\gamma}>0.7$ and $\pt>1.0\TeV$.
These events, which appear only in the observed data and were identified using a control region with the requirement $1.5 < \deta < 2.2$, arise from rare failures in certain ECAL reconstruction algorithms.
Second, in the later portion of the 2018 data-taking period, corresponding to 38.65\fbinv, a section of the HCAL was not active.
Because this effect was not included in the detector simulation, we reject events in which any narrow jet falls into the inactive section ($-3.05 < \eta < -1.35$, $-1.62 < \phi < -0.82$),
which reduces the signal efficiency by approximately 6\% within the affected data-taking period.
Third, the minimum angle between the jets and the \ptvecmiss is defined as
$\mindphi = \min[\Delta\phi(\vec{\widejet}_{1},\ptvecmiss),\,\Delta\phi(\vec{\widejet}_{2},\ptvecmiss)]$.
Events with $\mindphi < 0.8$, shown in Fig.~\ref{fig:nMinus1} (right), are selected because the missing momentum aligns with the jets in signal events.
The signal efficiency for the benchmark signal model after the final selection is 17\%.

We define two signal regions for the inclusive search, low-\RT ($0.15 < \RT \leq 0.25$) and high-\RT ($\RT > 0.25$).
Requiring $\RT > 0.25$ is found to maximize the signal sensitivity when only one signal region is considered.
However, including the low-\RT events as a separate signal region still reduces the expected cross section exclusion by an additional 60\% on average, and therefore we use both regions.
The low-\RT signal region also improves the sensitivity to signal models with $\rinv=0$,
which have smaller, but nonnegligible \ptmiss from decays of heavy flavor hadrons,
and therefore populate this region.

The values used in the requirements on \deta, \DRphispike, \mindphi, and \RT
are optimized to ensure maximal separation between signal and background,
using the figure of merit $F = \sqrt{2 \left[ \left(S+B\right) \, \ln \left(1+S/B\right) - S \right]}$~\cite{Cowan:2010js}.
$F$ is computed by comparing the yield $B$ of the simulated background processes to the yield $S$ of the signal model, considering events with \MT in the vicinity of the \PZprime boson mass.
For example, for the benchmark signal model with $\mZprime = 3.1\TeV$, events with $2.1 < \MT < 4.1\TeV$ are included in the yield computation.
The \mindphi requirement is optimized by restricting $B$ to the electroweak backgrounds;
the QCD background is omitted in this case because it is expected to have low values of \mindphi, much like the signal.
The requirements of the preselection and the final selection are summarized in Table~\ref{tab:sel}.

\begin{table}[!hbt]
\topcaption{Summary of the preselection and final selection requirements. The symbol * indicates a selection applied only to the later portion of the 2018 data.}
\centering
\begin{tabular}{M}
\multicolumn{2}{c}{Preselection requirements}\\
\hline
\pt(\widejet_{1,2}) > 200\GeV,&~\eta(\widejet_{1,2}) < 2.4 \\
\RT & >0.15 \\
\deta & <1.5 \\
\mt & >1.5\TeV \\
\nmuons & =0 \\
\neles & =0 \\
\ptmiss & \text{filters} \\
\DRphispike &> 0.1 \\
[2\cmsTabSkip]
\multicolumn{2}{c}{Final selection requirements}\\
\hline
\text{veto}~f_{\gamma}(\narrowjet_{1})>0.7 &\&~\pt(\narrowjet_{1}) > 1.0\TeV \\
\text{veto}~{-}3.05<\eta_{\narrowjet}<{-}1.35 &\&~{-}1.62<\phi_{\narrowjet}<{-}0.82~\text{*} \\
\mindphi & <0.8 \\
\end{tabular}
\label{tab:sel}
\end{table}

\section{Identification of semivisible jets}\label{sec:tagger}

The selection requirements described in Section~\ref{sec:evtsel} rely on event-level quantities or basic kinematic properties of the reconstructed jets, leptons, and missing transverse momentum.
Although they reject the vast majority of the SM background events,
the remaining background is still more than two orders of magnitude larger than the expected number of signal events for the benchmark signal model.
In order to reduce the background even further, we exploit the intrinsic differences between semivisible jets and SM jets.
A number of variables have already been derived in other contexts to characterize jets in terms of their substructure.
Many of these jet substructure variables may be used individually to provide weak discrimination between semivisible and SM jets.
To improve the discriminating power, the useful variables are combined in a BDT, whose output is an optimized discriminator
with values close to 1 for semivisible jets and close to 0 for SM jets.
Jets with a discriminator value higher than a chosen working point (WP) are considered to be ``tagged'' as semivisible.
Semivisible jet substructure~\cite{Park:2017rfb,Cohen:2020afv} and other tagging strategies~\cite{Bernreuther:2020vhm} have also been explored at the phenomenological level.

The 15 BDT input variables, computed for each jet, originate from several categories.
From heavy object identification, the $N$-subjettiness ratios \tauTO and \tauTT~\cite{Thaler:2010tr},
the energy correlation functions \Ntwo and \Nthree~\cite{Moult:2016cvt}, and the soft-drop mass \msd~\cite{Larkoski:2014wba} are used.
From quark-gluon discrimination, the jet girth \girth~\cite{Gallicchio:2012ez}, the major and minor jet axes \axmajor and \axminor~\cite{CMS-PAS-JME-13-002},
and the \pt dispersion \ptD~\cite{CMS-PAS-JME-13-002} are used.
The flavor-related variables used are the jet energy fractions for each type of constituent identified by the PF algorithm:
$f_{\Pchhad}$, $f_{\Pe}$, $f_{\Pgm}$, $f_{\Pnhad}$, and $f_{\gamma}$.
The angle between the jet and the missing transverse momentum, \dphij, is also included.
Variables sensitive to the multiplicity of jet constituents are deliberately excluded,
as this property is strongly correlated with the \rinv parameter of the signal model.

\begin{figure*}[htb!]
\centering
\includegraphics[width=0.49\linewidth]{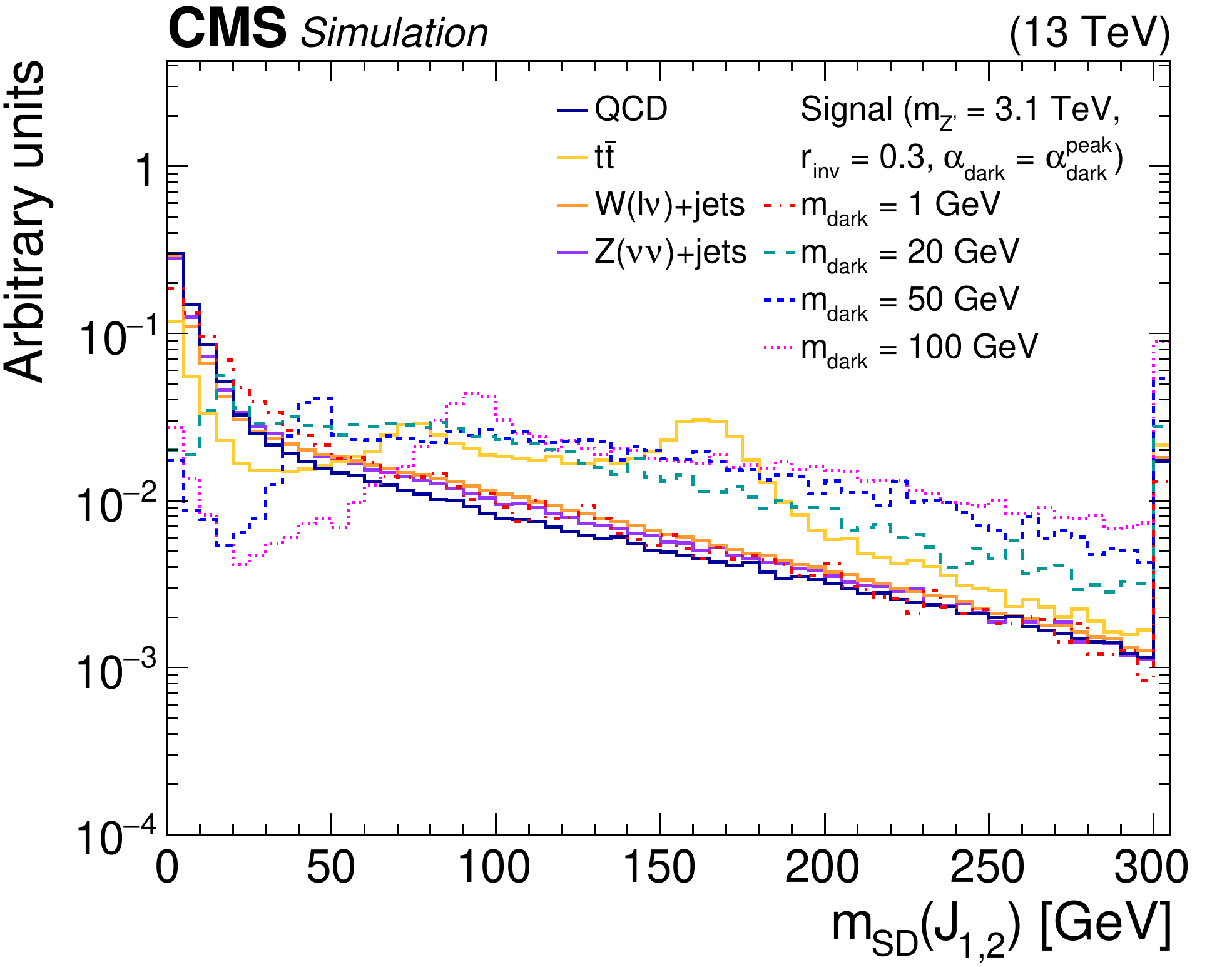}
\includegraphics[width=0.49\linewidth]{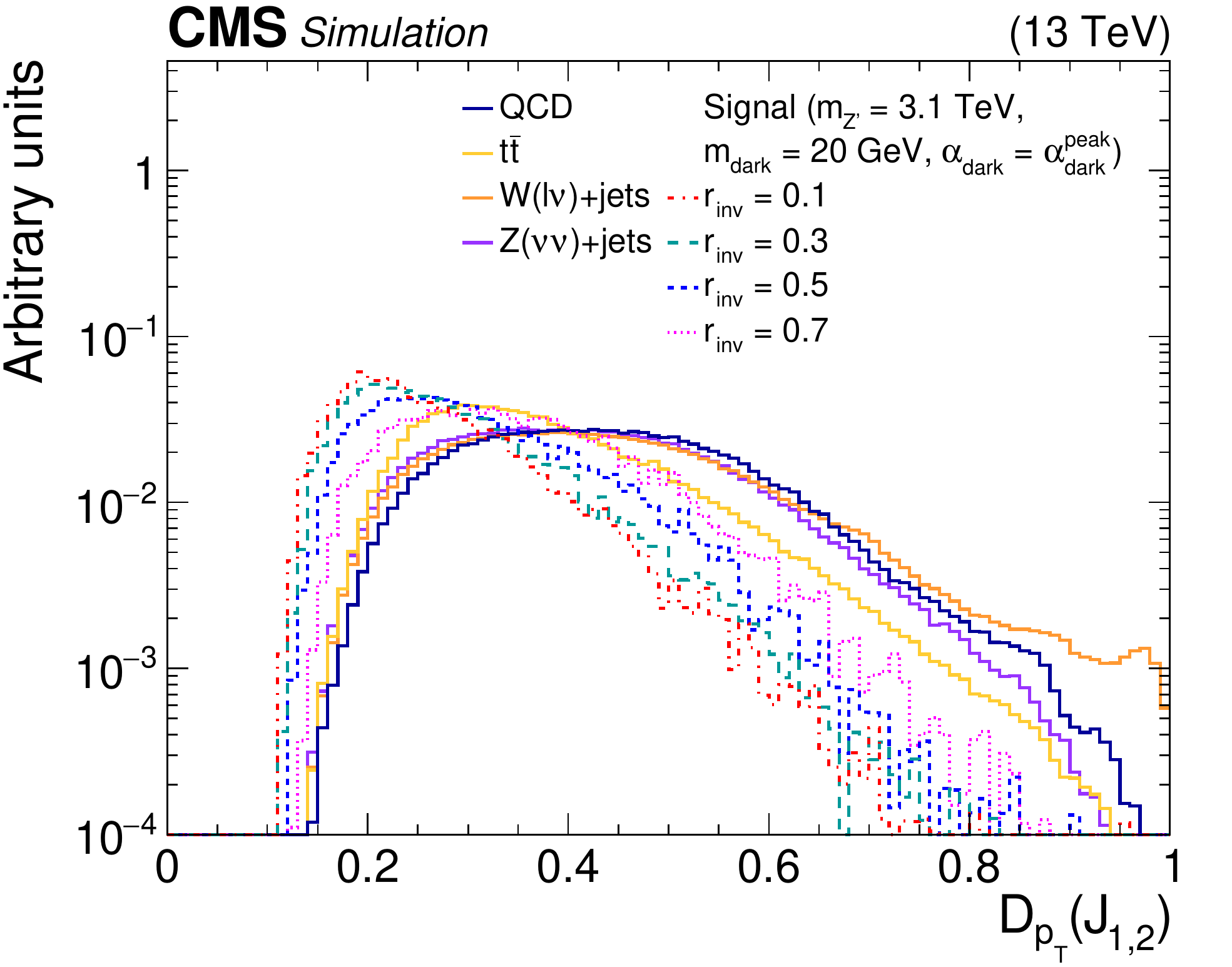}
\caption{The normalized distributions of the BDT input variables \msd and \ptD
for the two highest \pt jets from the simulated SM backgrounds and several signal models.
Each sample's jet \pt distribution is weighted to match a reference distribution (see text).
The last bin of each histogram includes the overflow events.
}
\label{fig:bdt_inputs}
\end{figure*}

As noted previously, semivisible jets are expected to have wider and softer radiation patterns compared to typical SM jets.
This is reflected in the quark-gluon variables, with the signal jets exhibiting larger values of \girth, \axmajor, and \axminor, and smaller values of \ptD.
In addition, the signal jets are very unlikely to have multiple prongs in their substructure,
which is reflected in their $N$-subjettiness and energy correlation function distributions.
Unlike boosted heavy objects such as top quarks or {\PW} or {\PZ} bosons, whose mass is reflected in the \msd of the resulting jet,
the mass of the originating dark quark is not reflected in the \msd distribution of semivisible jets.
This difference occurs because semivisible jets are formed by strong hadronization processes in both the dark and visible sectors,
rather than the two-prong weak decays of boosted heavy SM particles.
However, in semivisible jets, the \msd can reflect the dark hadron mass \mDark
when a single hard fragmentation carries a significant portion of the jet energy.
The jet energy fractions provide useful insight into the visible part of the signal jets,
and because they are normalized to the total visible jet energy, they induce very little dependence on the signal model parameters.
The distributions of \msd and \ptD are shown in Fig.~\ref{fig:bdt_inputs}
for the two highest \pt jets from the simulated SM backgrounds and several signal models,
with the jet \pt distributions weighted as described below.

The BDT is trained using the \textsc{scikit-learn} machine learning package~\cite{Pedregosa:2012toh} with the gradient boosting technique,
and additional diagnostic information is provided by the \textsc{rep} software~\cite{Likhomanenko:2015hca}.
The inputs are the two highest \pt jets from simulated signal and background samples,
with the variables described above computed for each jet.
We observe that the background simulation reproduces the observed distributions of the input variables.
Only jets from events that pass the preselection, defined in Section~\ref{sec:evtsel}, are considered.
The signal samples include various choices of parameters with relatively high acceptance for the preselection:
$\mZprime \geq 1.5\TeV$, $\mDark \geq 10\GeV$, $0.1 \leq \rinv \leq 0.8$, and all \aDark variations.
Each signal sample is weighted equally.
Jets are classified as signal jets if they originate from one of the signal samples
and if they are spatially associated with generator-level dark sector particles.
A weight is assigned to each jet so that the \pt distributions of the signal and background jets match that of a benchmark signal with $\mZprime = 3.0\TeV$.
This prevents the BDT from learning differences in the \pt distribution that arise from the originating process rather than the intrinsic nature of the jets,
and it also prevents the BDT from forcing the background \MT distribution to resemble that of the signal.
The background jets are taken from an equal mix of the QCD and \ttbar processes.
It was found that training on only one of those two background processes caused the BDT to misclassify jets from the other background process as signal jets at a rate of 10--20\%.
The BDT hyperparameters are optimized to maximize the signal efficiency at a background efficiency of 10\%,
while requiring low values of variables that measure overtraining and the correlation of the background efficiency with \MT.

\begin{figure*}[htb!]
\centering
\includegraphics[width=0.49\linewidth]{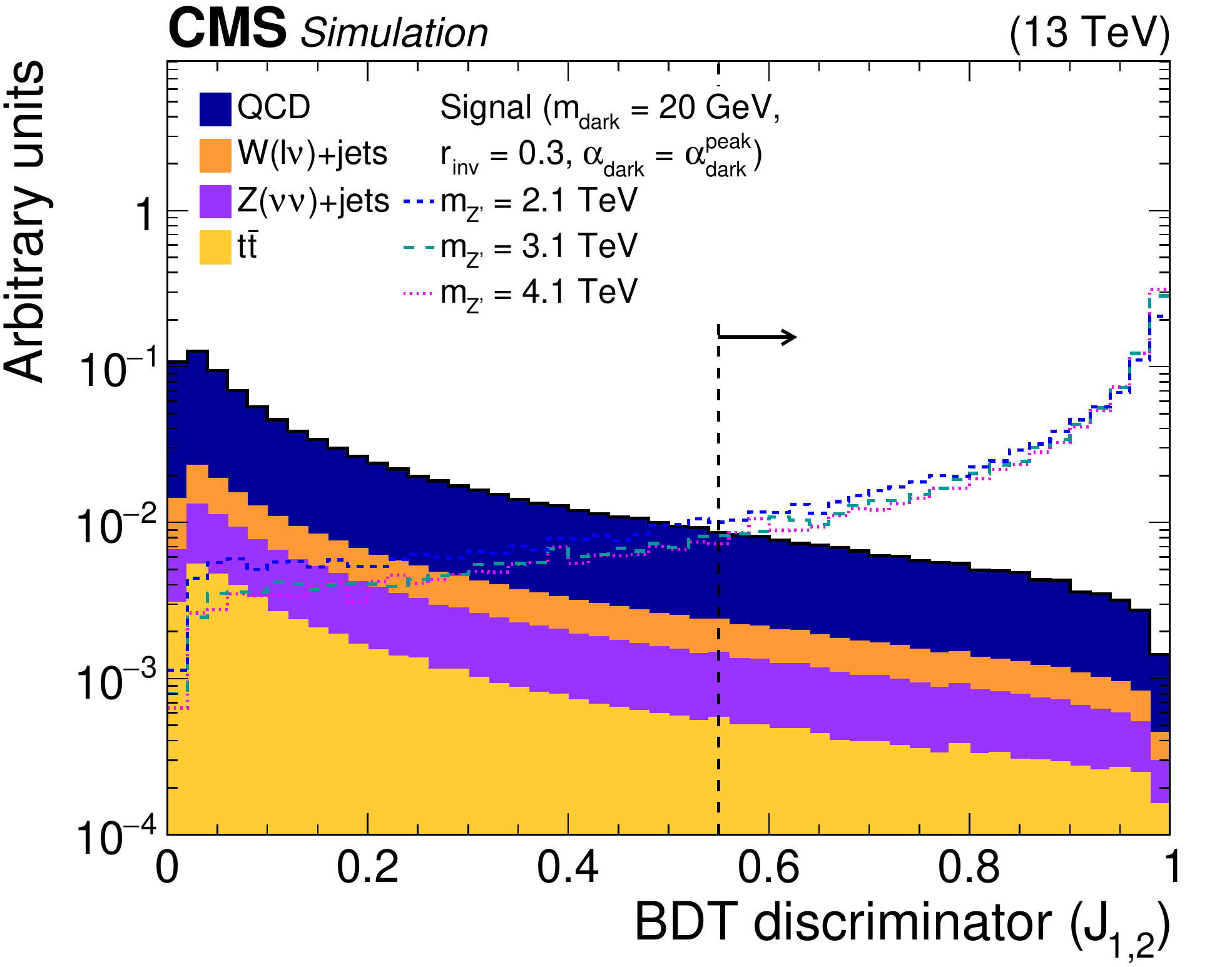}
\includegraphics[width=0.49\linewidth]{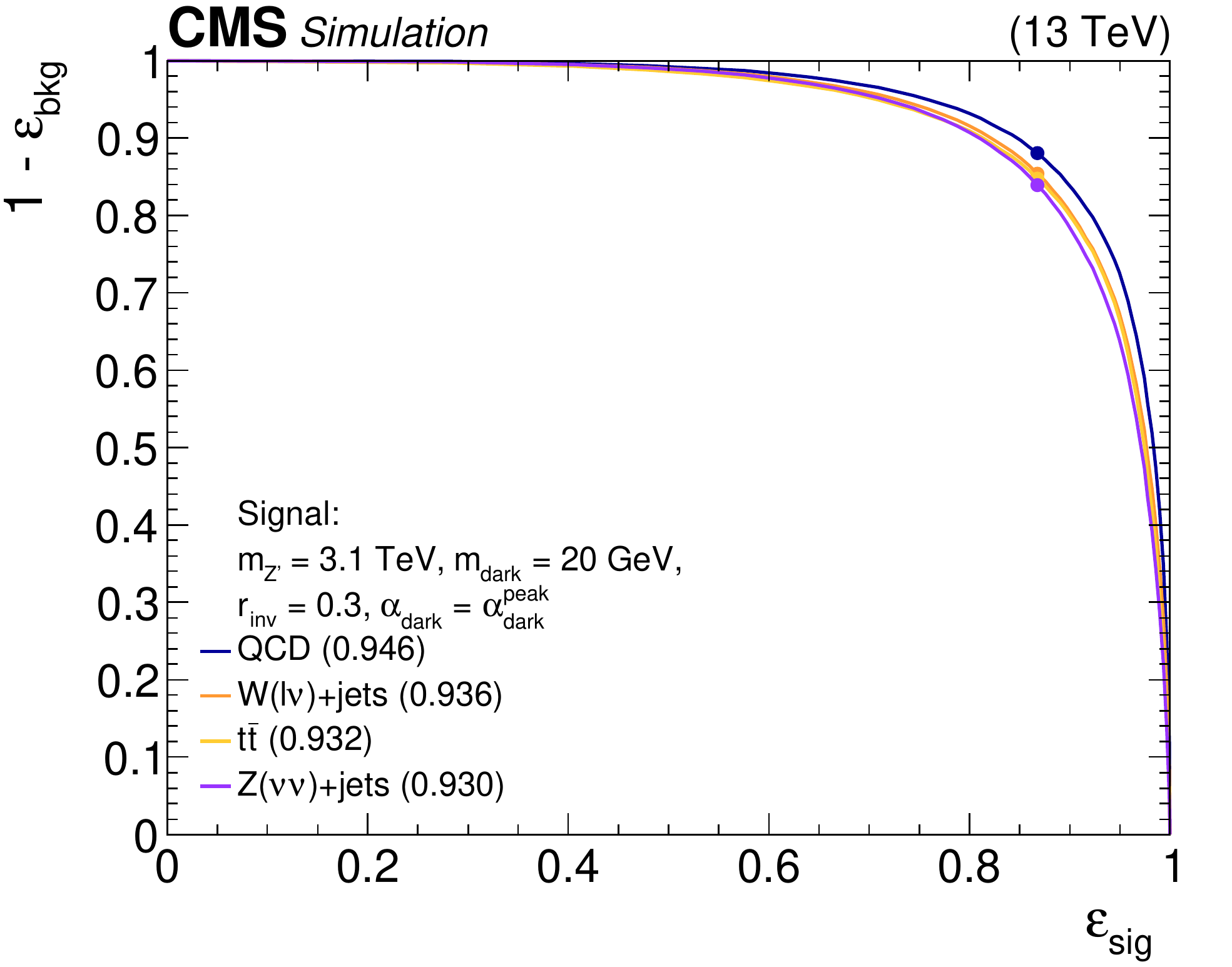}
\caption{Left: The normalized BDT discriminator distribution for the two highest \pt jets from the simulated SM backgrounds and several signal models.
The discriminator WP of 0.55 is indicated as a dashed line.
Right: The BDT ROC curves for the two highest \pt jets, comparing the simulated SM backgrounds with one signal model.
The area under the ROC curve is listed in parentheses for each pairing.
The discriminator WP of 0.55 is indicated on each curve as a filled circle.
}
\label{fig:roc}
\end{figure*}

The BDT exhibits strong and uniform rejection of jets from all of the SM background processes, while preserving a significant amount of the signal.
The discriminator distribution and the receiver operator characteristic (ROC) curve are shown in Fig.~\ref{fig:roc}.
The five most important variables in the BDT are found to be the \msd, \dphij, \tauTO, \tauTT, and \ptD.
Jet tagging algorithms are typically evaluated using three metrics:
the accuracy (Acc.) of classifying signal jets (at a working point of 0.5), the area under the ROC curve (AUC),
and the background rejection (\bkgrej) at a signal efficiency (\sigeff) of 30\%.
These metrics are provided for each background process, compared to the benchmark signal process, in Table~\ref{tab:bdt_metrics}.
The signal efficiency of the BDT exhibits some additional dependence on \mDark for several reasons.
First, the important variable \msd is less useful to distinguish signal jets at very low \mDark, where they more closely resemble QCD,
and at very high \mDark, where they more closely resemble boosted \PW boson jets from \ttbar events.
Second, at very low \mDark, unstable dark hadrons cannot decay to \PQb or \PQc quarks, which changes the flavor content of the jets,
including the jet energy fractions used in the BDT.

The final WP value is chosen to be 0.55; at this WP, the BDT rejects 84--88\% of simulated background jets,
while correctly classifying 87\% of jets from the benchmark signal model.
This choice retains sufficient events to facilitate the background estimation, which will be described in Section~\ref{sec:bkg}.
When the BDT is employed, we select subsets of the high-\RT and low-\RT inclusive signal regions
by requiring that both jets in each event are tagged as semivisible.
These signal regions are called high-SVJ2 and low-SVJ2.
Events in which only one jet is tagged as semivisible are not found to provide significant additional sensitivity.
Because semivisible jets do not occur in the SM,
there is no viable approach to assess any difference in the tagger performance between data and simulation in signal events.
Therefore, the results from the BDT-based regions assume
that any residual differences are covered by the various other systematic uncertainties in the signal,
which are described in Section~\ref{sec:syst}.

\begin{table}[htb]
\topcaption{Metrics representing the performance of the BDT for the benchmark signal model ($\mZprime = 3.1\TeV$, $\mDark = 20\GeV$, $\rinv = 0.3$, $\aDark = \aDarkPeak$),
compared to each of the major SM background processes.}
\centering
\begin{tabular}{lrrr}
Background & Acc. & AUC & \bkgrej (\sigeff = 0.3) \\
\hline
QCD    & 0.883 & 0.946 & 636.8\colspace \\
\ttbar & 0.883 & 0.932 & 290.0\colspace \\
\wjets & 0.883 & 0.936 & 477.1\colspace \\
\zjets & 0.883 & 0.930 & 455.4\colspace \\
\end{tabular}
\label{tab:bdt_metrics}
\end{table}

\section{Background estimation}\label{sec:bkg}

As described in the previous sections, QCD is the dominant SM background process,
while the electroweak processes \ttbar, \wjets, and \zjets provide smaller but nonnegligible contributions.
The proportion of each background varies depending on the signal region,
with QCD comprising 83--88\% in the low-\RT and low-SVJ2 regions,
but only 46--57\% in the high-\RT and high-SVJ2 regions.
In each region, the \MT distribution from each SM background processes is expected to be smoothly falling.
The observed \MT distributions from each data-taking period are added together in each signal region,
producing summed distributions that are used for the final results.

We fit the observed \MT distribution in each signal region with a functional form that represents this expected behavior,
following the approach from traditional dijet searches, such as Ref.~\cite{Sirunyan:2019vgj}.
The primary fit function is:
\begin{equation}
g(x) = \exp(p_1 x) x^{p_2 [1 + p_3 \ln(x)]},\label{eq:bkgfn}
\end{equation}
where $x = \MT/\!\sqrts$ (with $\sqrts=13\TeV$) and $p_{1}$, $p_{2}$, and $p_{3}$ are free parameters in the fit.
The specific form of this function is chosen to improve numerical stability and reduce parameter correlations, in order to ensure a stable fit.
To prevent the minimizer from finding false minima in the parameter space of the fit function,
we adopt an approach in which many combinations of initial values, varying both sign and magnitude, are tested for each parameter,
and the resulting fit with the lowest $\chi^2$ value is chosen.
Versions of the function with different numbers of free parameters are compared to each other,
and the optimal number of parameters for the fit in each signal region is determined using the Fisher $F$-test~\cite{Lomax:2012xyz}.
The high-\RT region requires all three parameters,
while the other three signal regions use the two-parameter version of the function, with $p_3$ set to 0.

An analytic fit could produce a biased background prediction if it were not sufficiently flexible to adapt to observed data that actually followed a different distribution.
To quantify this potential bias, we also fit the data in each signal region with two secondary functions, one similar to those used in Refs.~\cite{Cohen:2015toa,Sirunyan:2019vgj}
and another based on the family used in Refs.~\cite{UA2:1990gao,UA2:1993tee}.
We then randomly generate artificial pseudodata sets distributed according to the secondary function fits, with a similar yield to the observed data.
For each pseudodata set, a maximum likelihood fit is performed, including both the background prediction from the primary function and a signal model with varying cross section.
When generating these data sets, two cases are considered: no signal events injected, or signal events injected at the nominal \PZprime cross section.
The method of varying initial values is also applied here, to ensure the best possible fit of the primary function to the secondary function.
The result of each likelihood fit is an extracted cross section, which is compared to the injected cross section in the figure of merit $b = (\rext-\rinj)/\rerr$,
where \rerr is the estimated uncertainty in the extracted cross section.
It is found that $\abs{b} \leq 0.5$ for both secondary functions and both signal injection cases.
The background prediction is therefore determined to be sufficiently unbiased.

\begin{figure*}[htb!]
\centering
\includegraphics[width=0.49\linewidth]{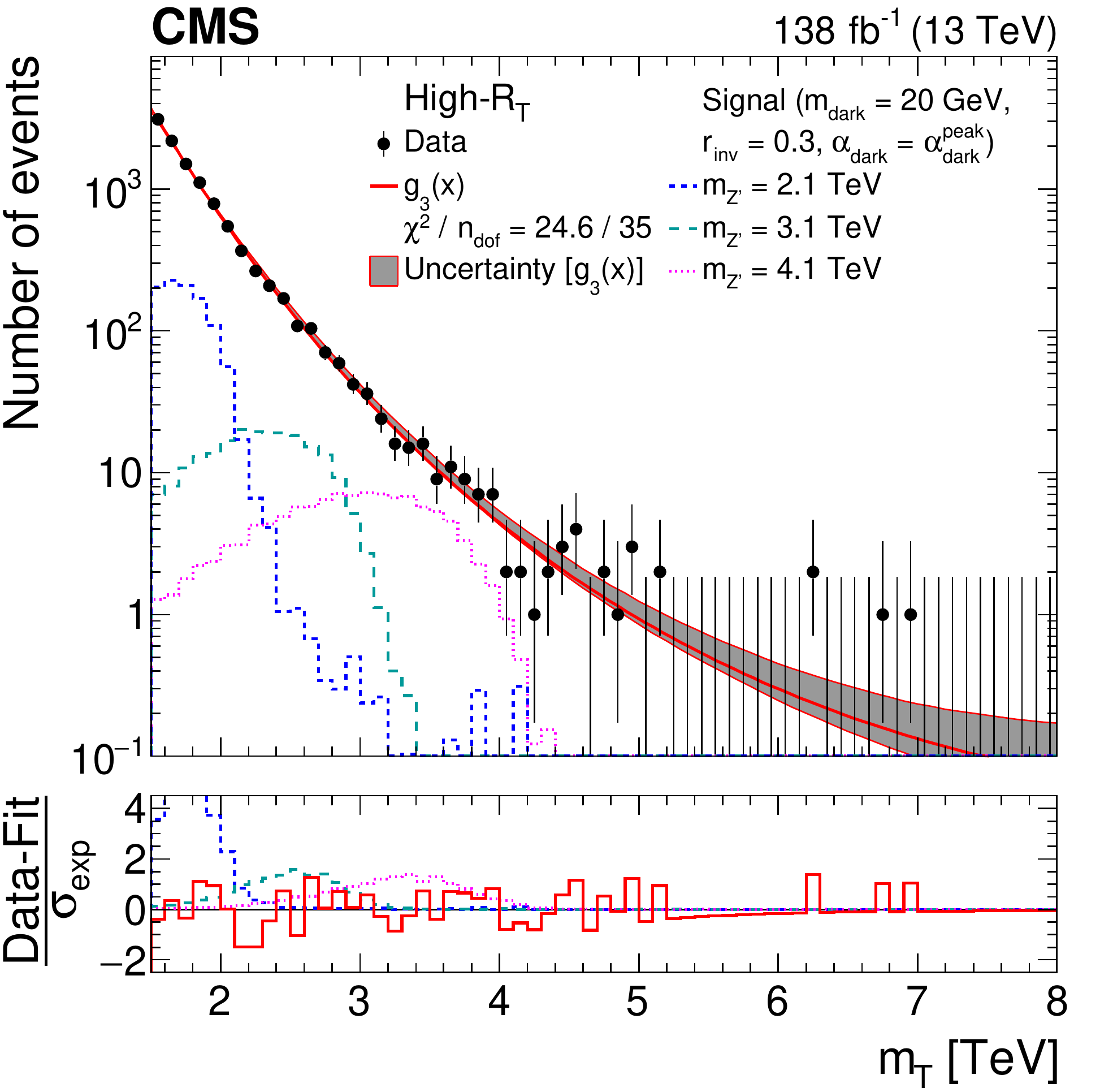}
\includegraphics[width=0.49\linewidth]{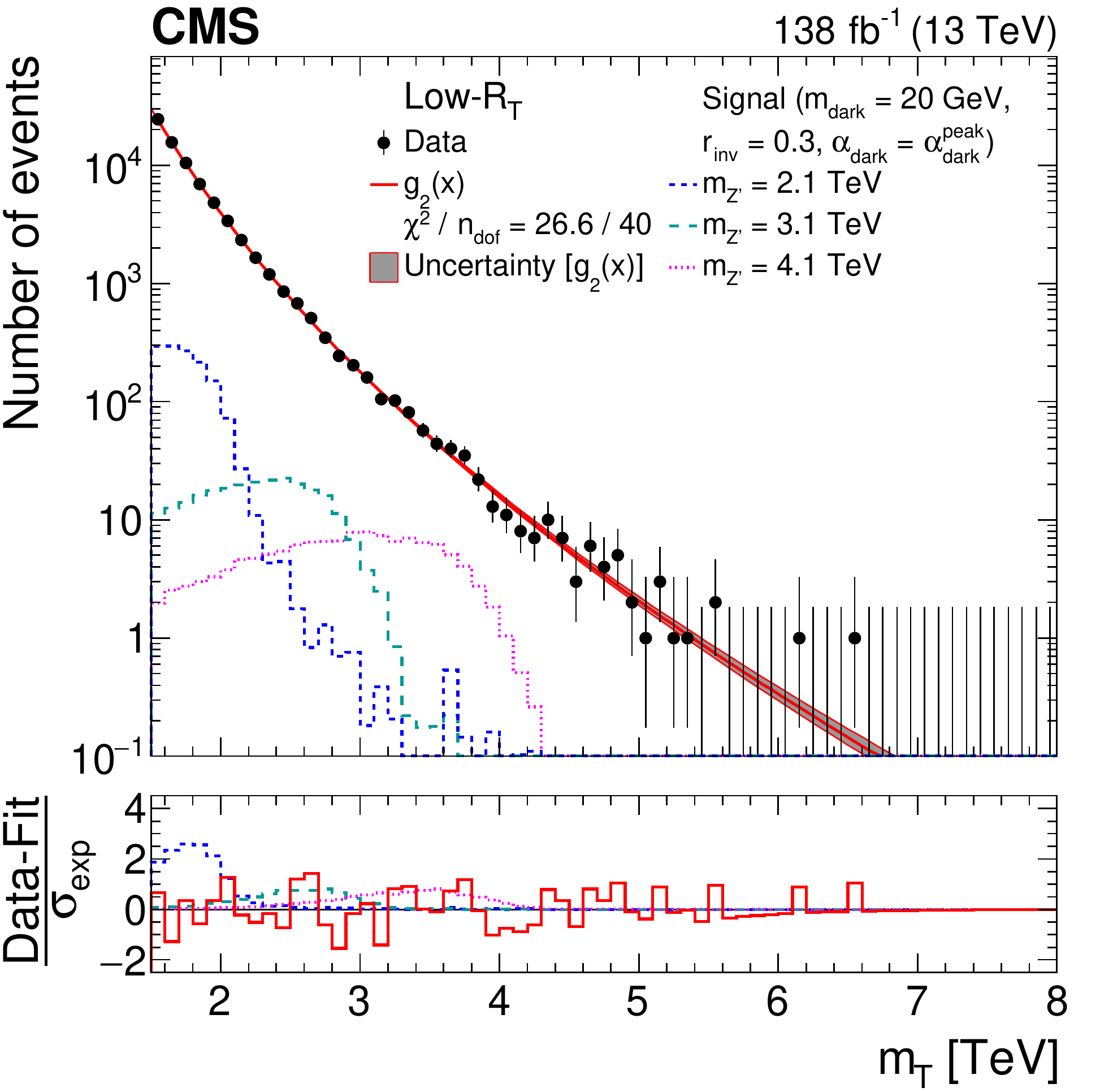}
\caption{The \MT distribution for the high-\RT (left) and low-\RT (right) signal regions,
comparing the observed data to the background prediction from the analytic fit ($g_{3}(x) = \exp(p_1 x) x^{p_2 [1 + p_3 \ln(x)]}$, $g_{2}(x) = \exp(p_1 x) x^{p_2}$, $x = \MT/\!\sqrt{s}$).
The lower panel shows the difference between the observation and the prediction
divided by the statistical uncertainty in the observation (\sigmaexp).
The distributions from several example signal models, with cross sections corresponding to the observed limits, are superimposed.
}
\label{fig:prefit_cut}
\end{figure*}

\begin{figure*}[htb!]
\centering
\includegraphics[width=0.49\linewidth]{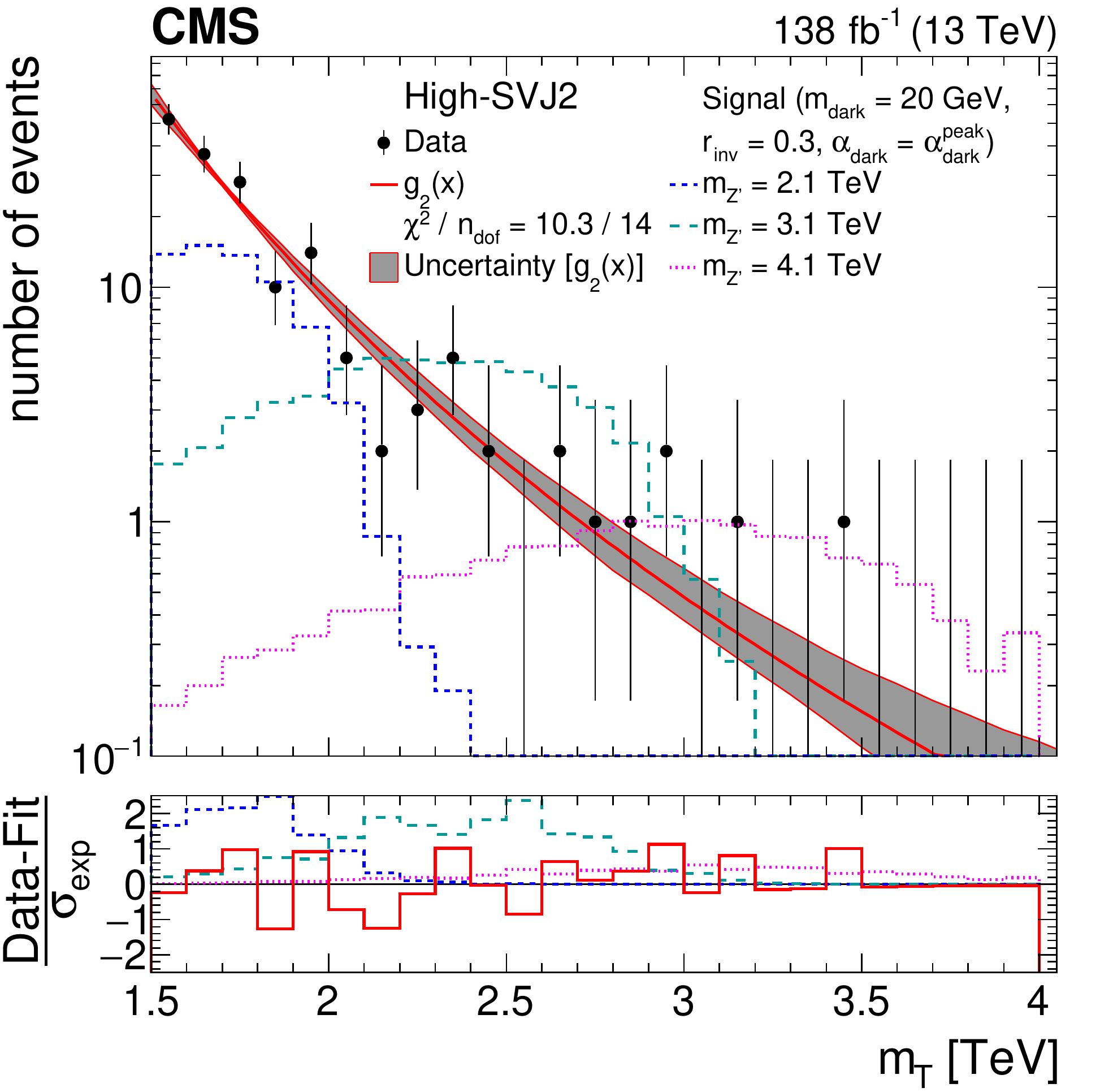}
\includegraphics[width=0.49\linewidth]{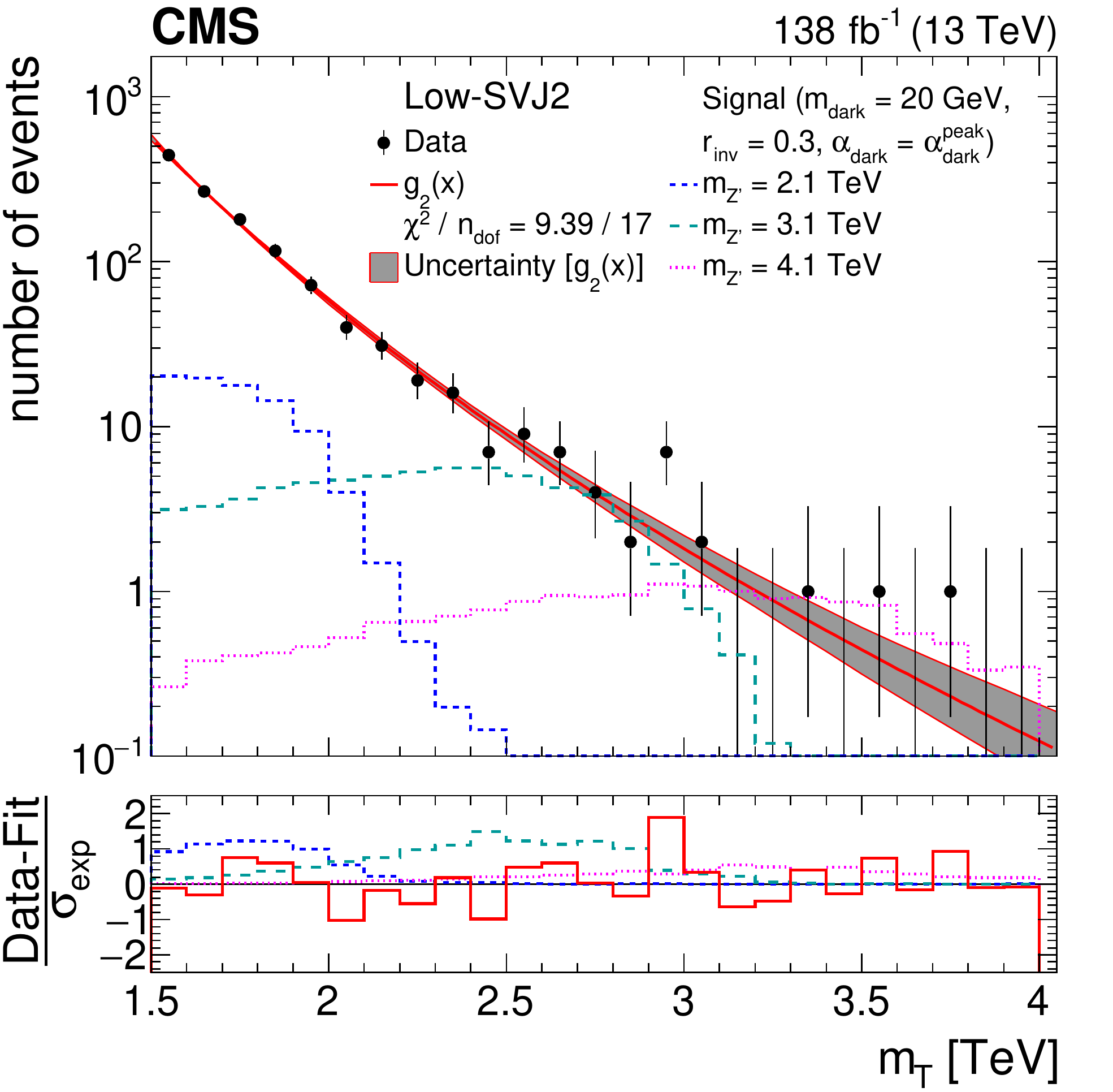}
\caption{The \MT distribution for the high-SVJ2 (left) and low-SVJ2 (right) signal regions,
comparing the observed data to the background prediction from the analytic fit ($g_{2}(x) = \exp(p_1 x) x^{p_2}$, $x = \MT/\!\sqrt{s}$).
The lower panel shows the difference between the observation and the prediction
divided by the statistical uncertainty in the observation (\sigmaexp).
The distributions from several example signal models, with cross sections corresponding to the observed limits, are superimposed.
}
\label{fig:prefit_bdt}
\end{figure*}

The results of the background-only fits are compared to the observed data in Fig.~\ref{fig:prefit_cut} for the inclusive signal regions
and in Fig.~\ref{fig:prefit_bdt} for the BDT-based regions.
Applying the BDT to tag semivisible jets reduces the background by almost two orders of magnitude.
A new resonance would appear as a significant excess of events in a range of \mt values, and no such excess with respect to the predictions is observed.
The few events at high \mt values in the high-\RT region are consistent with the background prediction, within uncertainties.
For $5 < \mt < 8\TeV$, we observe 6 events, while integrating the background-only fit predicts $8.4^{{+}2.1}_{{-}1.4}$;
for $6 < \mt < 7\TeV$, we observe 4 events, while the prediction is $2.0^{{+}0.6}_{{-}0.4}$.
We proceed to set limits on the effective cross sections for different signal models in Section~\ref{sec:results}.

\section{Systematic uncertainties}\label{sec:syst}

Various systematic uncertainties in the signal simulation are assessed.
Most uncertainties arise from one of two possible sources.
First, experimental uncertainties relate to measurements and detector and reconstruction effects, often occurring when a correction is applied to handle some difference between data and simulation.
Second, theory uncertainties reflect possible variations in the generation of signal events.
Some uncertainties, called ``flat'' uncertainties, affect just the signal yield, while other uncertainties, called ``shape'' uncertainties, affect both the yield and distribution of signal events.
Experimental shape uncertainties are considered to be uncorrelated between each year of data taking.
Uncertainties that affect just the yield and all theory uncertainties are considered to be correlated between each year of data taking,
because their sources are common across all years.
Finally, the statistical uncertainty from the limited number of simulated signal events is considered in each bin of the \MT distributions.
The statistical uncertainty is uncorrelated for each bin of the \MT distribution and for each year of data taking.
The effect of each uncertainty on the signal yield is summarized in Table~\ref{tab:syst}.

The flat experimental uncertainties include a 1.6\% uncertainty in the measurement of the integrated luminosity of the data~\cite{CMS:2021xjt,CMS-PAS-LUM-17-004,CMS-PAS-LUM-18-002}
and a 2.0\% uncertainty in the measured trigger efficiency to account for potential kinematic differences between the signal regions and the control region used in the measurement.
The remaining experimental uncertainties are shape uncertainties.
These include uncertainties in the jet energy corrections and the jet energy resolution, which are evaluated depending on the jet \pt and $\eta$,
and propagated to all jet-related kinematic variables and to the missing transverse momentum.
The uncertainty in the pileup reweighting arises from the variation of the total inelastic cross section by 5\%~\cite{Sirunyan:2018nqx}.
The statistical uncertainty in the muon trigger control region is propagated to the trigger efficiency correction as an additional systematic uncertainty.

The theory uncertainties are generally treated as shape uncertainties.
The uncertainty in the PDFs is assessed by reweighting the generated events using the different PDF replicas~\cite{Ball:2013hta,Ball:2017nwa};
only the effect on the acceptance is considered.
The uncertainty in the renormalization and factorization scales is computed by varying each scale by a factor of two~\cite{Kalogeropoulos:2018cke,Catani:2003zt,Cacciari:2003fi};
as with the PDF uncertainty, only the effect on the acceptance is considered.
The uncertainty in the parton shower model is found by similar variations of the renormalization scale used in \PYTHIA~\cite{Mrenna:2016sih};
the contributions to initial-state radiation (ISR) and final-state radiation (FSR) are considered separately.
Because the jet energy corrections are measured for SM jets, whose constituents may be different than semivisible jets,
we include an additional uncertainty in the jet energy scale, which is derived by comparing the generator-level jet \pt to the reconstructed jet \pt.
The magnitude of the jet energy scale variation is typically 5\%, in contrast with the 2\% effect found when the same procedure is applied to the simulated electroweak backgrounds.
As with the jet uncertainties described above, this effect is propagated to all jet-related variables and to the missing transverse momentum.

\begin{table}[!hbtp]
\topcaption{The range of effects on the signal yield for each systematic uncertainty and the total. Values less than 0.01\% are rounded to 0.0\%.}
\centering
\begin{tabular}{lr@{\hspace{0em}}l}

Uncertainty & Yield ef&fect [\%] \\
\hline
Integrated luminosity & 1&.6 \\
Jet energy corrections & 0.05&--12 \\
Jet energy resolution & 0.02&--2.3 \\
Jet energy scale & 0.29&--21 \\
PDF & 0.0&--5.3 \\
Parton shower FSR & 0.07&--9.4 \\
Parton shower ISR & 0.0&--2.9 \\
Pileup reweighting & 0.0&--1.3 \\
Renormalization and factorization scales & 0.0&--0.12 \\
Statistical & 1.2&--4.9 \\
Trigger control region & 0.24&--0.40 \\
Trigger efficiency & 2&.0 \\
[\cmsTabSkip]
Total & 3.3&--22 \\
\end{tabular}
\label{tab:syst}
\end{table}

The parameters of the background fit function in each region, including the normalization, are freely floating.
The uncertainties in these parameters arise from the statistical uncertainty in the data.
As mentioned in Section~\ref{sec:bkg}, the background prediction is found to be sufficiently unbiased that no additional uncertainty is needed.
Overall, the largest impact on the sensitivity of the search comes from the background normalization in the high-\RT signal region, which can change by up to 10\%.
This is found to reduce the sensitivity, in terms of the excluded cross section, by a factor of roughly 2--4.

\section{Results}\label{sec:results}

When interpreting the results, we consider the two inclusive signal regions together, and separately consider the two BDT-based signal regions together.
The combined likelihood for each pair of signal regions is computed as the product of the likelihood from each bin in the \MT distributions.
We set limits on the effective cross section $\sigmaZprime\Bdark$ at 95\% confidence level (\CL) using the modified frequentist \CLs approach~\cite{Junk:1999kv,Read:2002hq}.
The systematic uncertainties described above are included in the maximum likelihood fit as nuisance parameters,
with the flat uncertainties given log-normal prior distributions and the shape uncertainties given Gaussian prior distributions.
This interpretation procedure is tested and validated by repeating the bias testing procedure described in Section~\ref{sec:bkg} for each pair of signal regions.
The background fit parameters for each signal region are still treated as independent,
but the injected and extracted signal cross sections are constrained to be the same in both regions in the pair.
As in Section~\ref{sec:bkg}, we find that the procedure is sufficiently unbiased.

\begin{figure*}[htb!]
\centering
\includegraphics[width=0.49\linewidth]{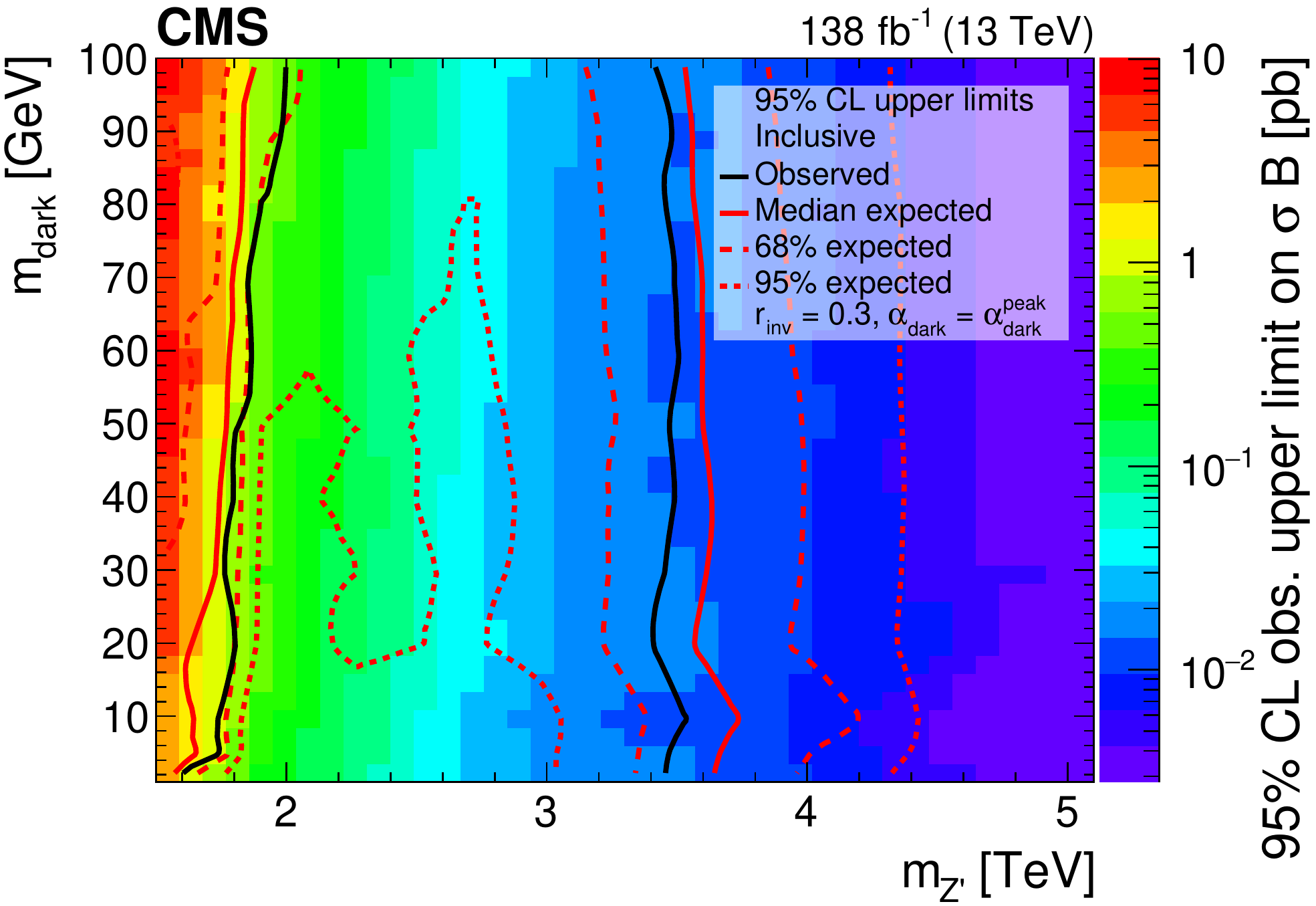}
\includegraphics[width=0.49\linewidth]{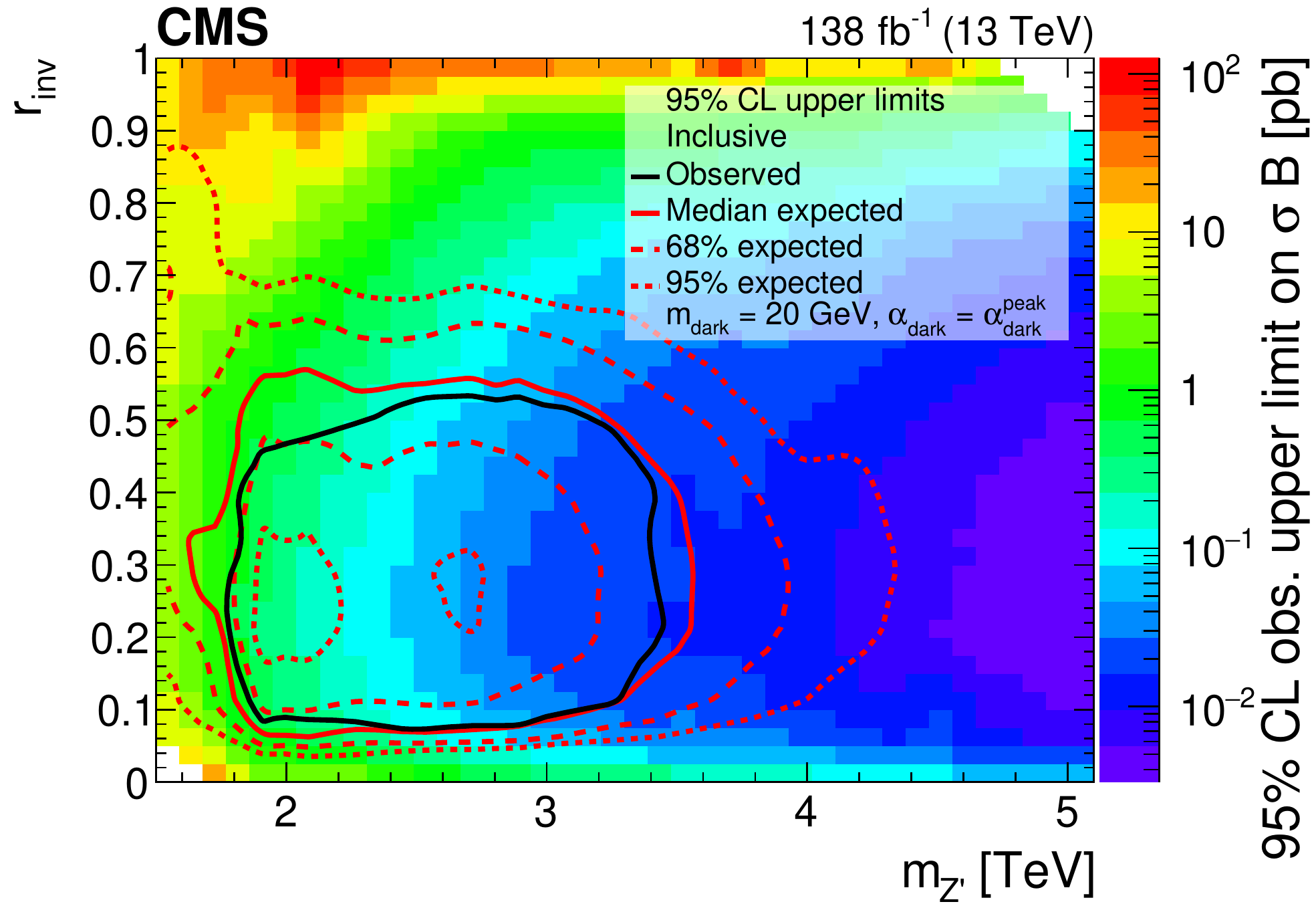}\\
\includegraphics[width=0.49\linewidth]{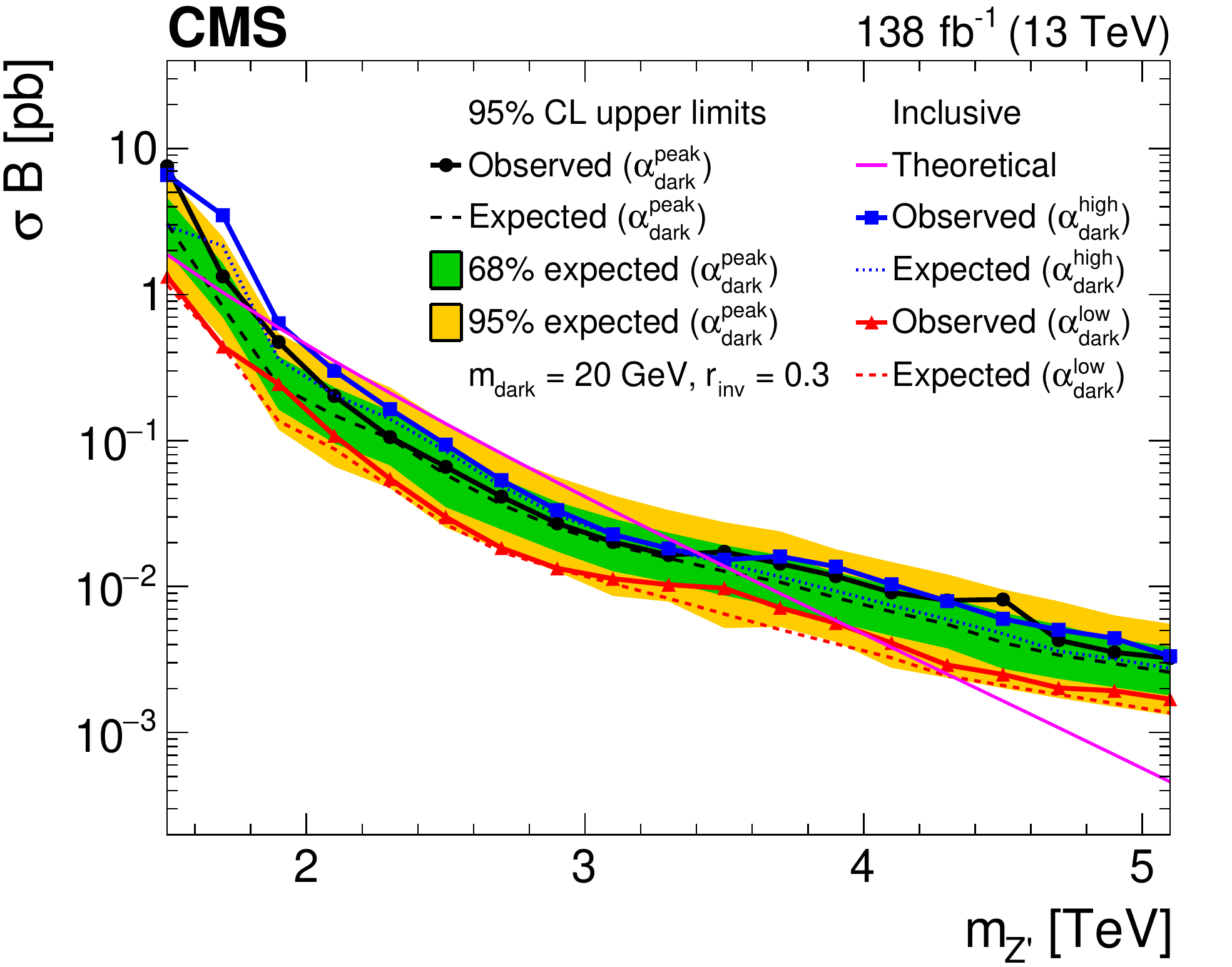}
\caption{The 95\% \CL upper limits on the product of the cross section and branching fraction from the inclusive search, for variations of pairs of the signal model parameters.
In the upper plots, the filled region indicates the observed upper limit. The solid black curves indicate the observed exclusions for the nominal \PZprime cross section,
while the solid red curves indicate the expected exclusions, and the dashed lines indicate the regions containing 68 and 95\% of the distributions of expected exclusions.
In the upper left plot, the regions between the respective pairs of lines or below the inner 95\% dashed line are excluded.
In the upper right plot, the regions inside the circles are excluded.
The lower plot shows the \aDark variation as multiple curves in one dimension, because there are only three parameter values considered.
The black, blue, and red solid lines show the observed upper limits for each variation, while the dashed lines show the expected limits.
The inner (green) and outer (yellow) bands indicate the regions containing 68 and 95\%, respectively, of the distributions of expected limits.
The purple solid line labeled ``Theoretical'' represents the nominal \PZprime cross section.
}
\label{fig:limit_cut_2D}
\end{figure*}

\begin{figure*}[htb!]
\centering
\includegraphics[width=0.49\linewidth]{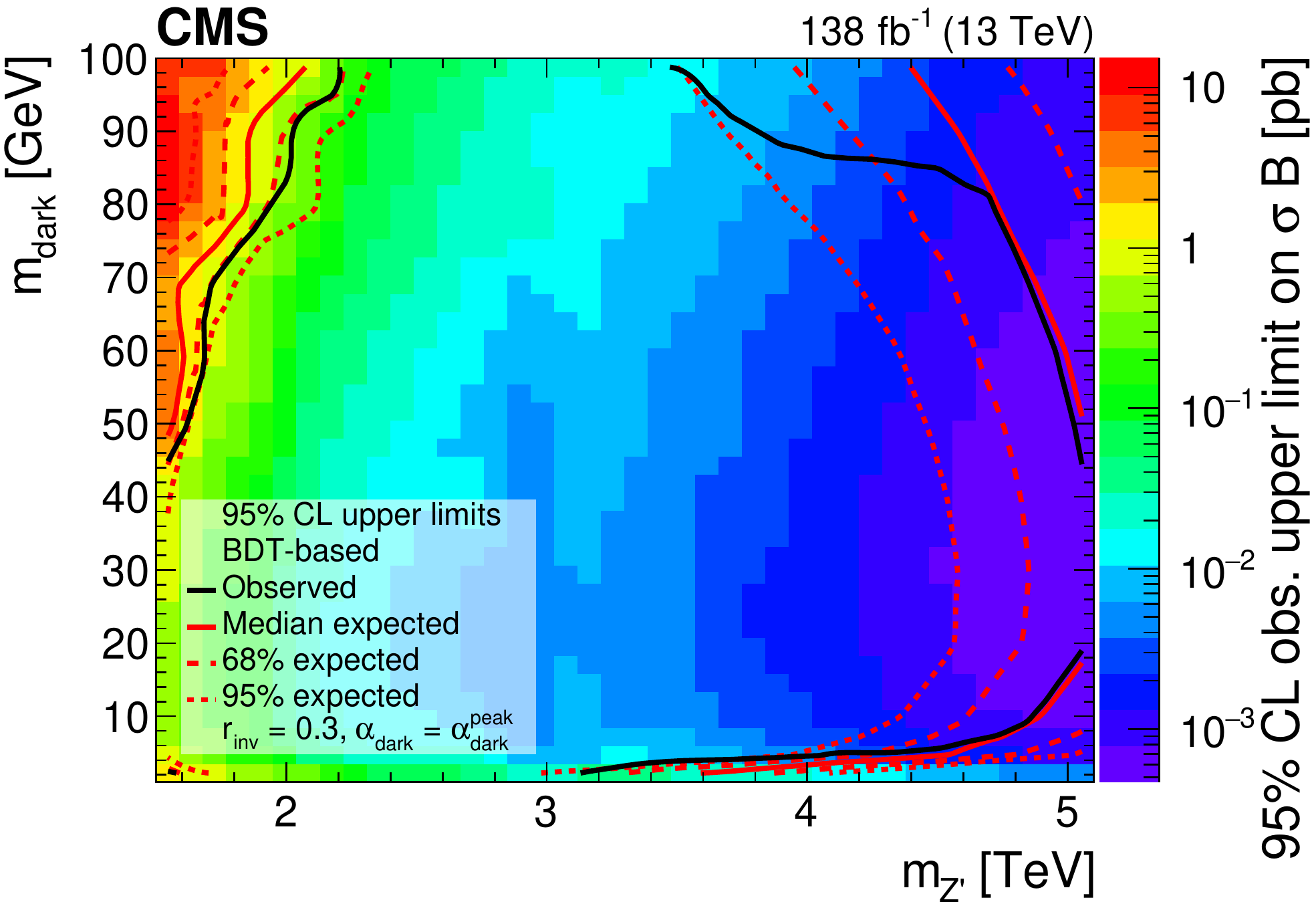}
\includegraphics[width=0.49\linewidth]{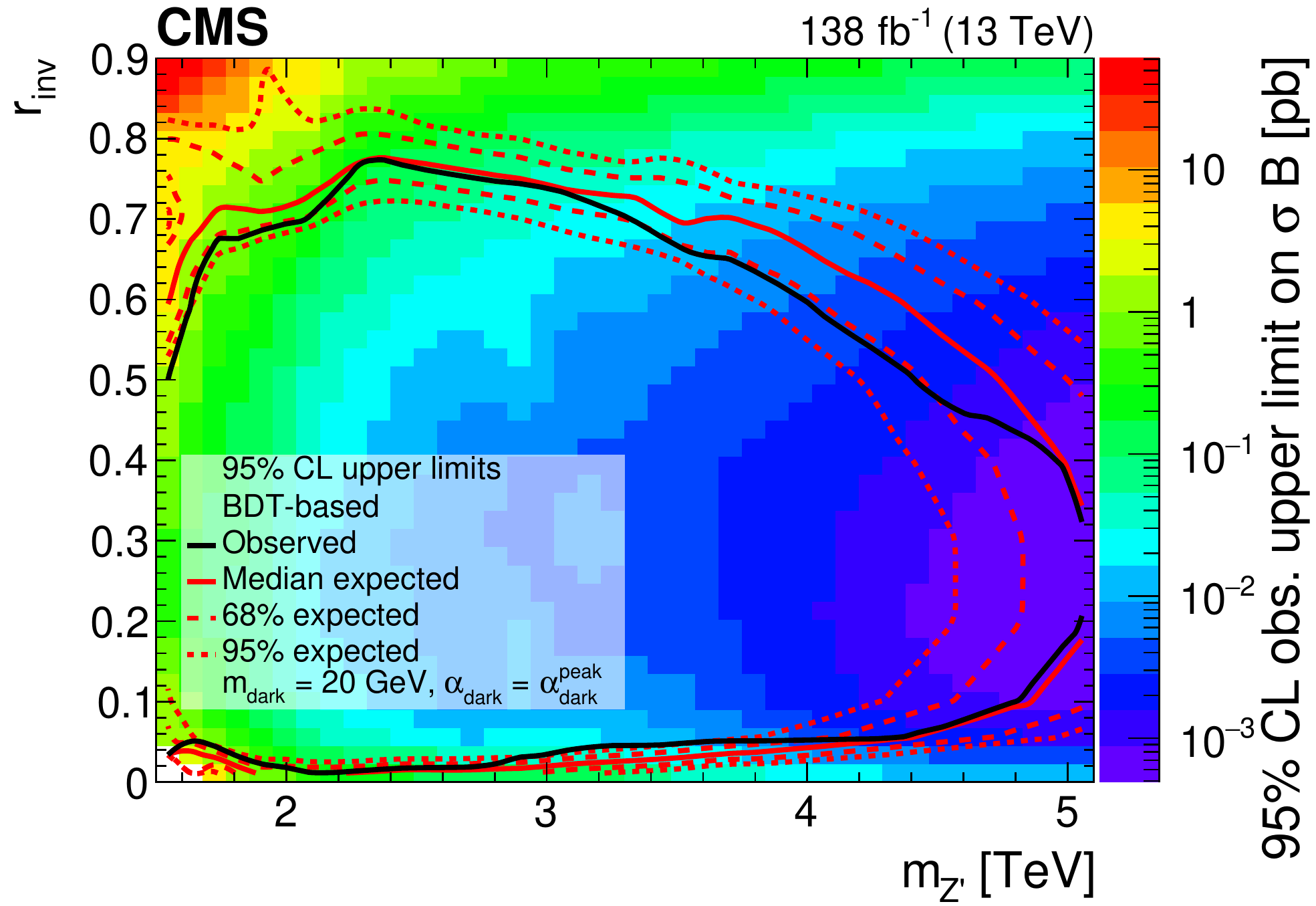}\\
\includegraphics[width=0.49\linewidth]{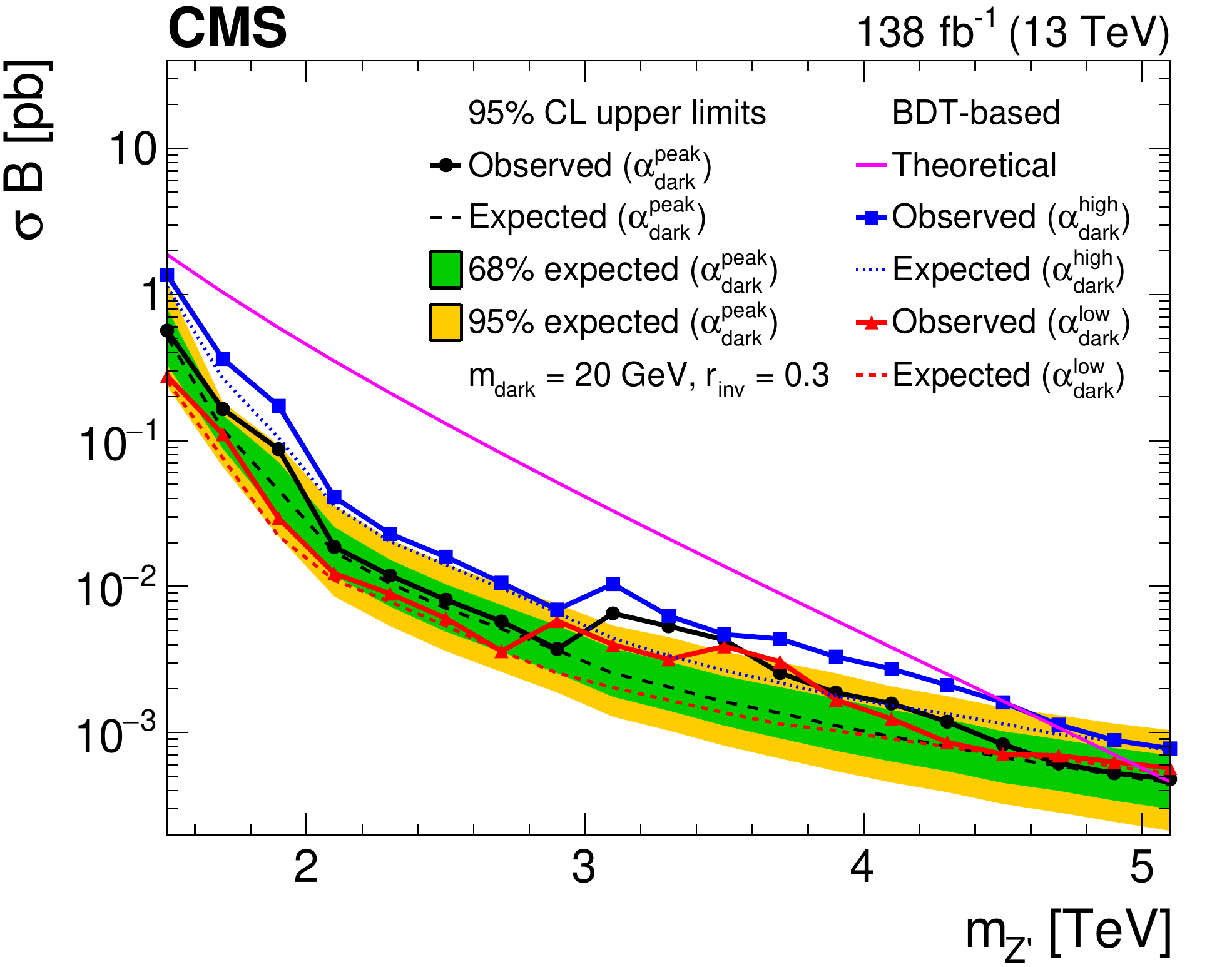}
\caption{The 95\% \CL upper limits on the product of the cross section and branching fraction from the BDT-based search, for variations of pairs of the signal model parameters.
In the upper plots, the filled region indicates the observed upper limit. The solid black curves indicate the observed exclusions for the nominal \PZprime cross section,
while the solid red curves indicate the expected exclusions, and the dashed lines indicate the regions containing 68 and 95\% of the distributions of expected exclusions.
The regions inside the circles are excluded.
The lower plot shows the \aDark variation as multiple curves in one dimension, because there are only three parameter values considered.
The black, blue, and red solid lines show the observed upper limits for each variation, while the dashed lines show the expected limits.
The inner (green) and outer (yellow) bands indicate the regions containing 68 and 95\%, respectively, of the distributions of expected limits.
The purple solid line labeled ``Theoretical'' represents the nominal \PZprime cross section.
}
\label{fig:limit_bdt_2D}
\end{figure*}

The expected limits are derived from a special data set created by replacing the observed data with the central values of the background predictions from the analytic fits
when comparing the likelihoods of the background-only and signal-plus-background hypotheses.
The likelihood distributions from the special data set and the observed data are computed as a function of the effective cross section
and are used to calculate the \CLs criterion following the asymptotic approximation~\cite{Cowan:2010js}.
The expected and observed limits are shown in two-dimensional planes of the signal parameters for the inclusive search in Fig.~\ref{fig:limit_cut_2D} and for the BDT-based search in Fig.~\ref{fig:limit_bdt_2D}.
The exclusion contours are defined by comparison to the product of the theoretical cross section for \PZprime boson production and the \PZprime branching fraction to dark quarks.
The results may be reinterpreted for different values of the \PZprime couplings and branching fractions in the range of validity of the narrow width approximation, ${\wZprime/\mZprime \lesssim 10\%}$~\cite{Boveia:2016mrp}.
To provide a continuous distribution, the limits for simulated signal models are logarithmically interpolated.

The results from the inclusive signal regions exclude observed (expected) values of up to $1.5 < \mZprime < 4.0\TeV$ ($1.5 < \mZprime < 4.3\TeV$), with the widest exclusion range for models with $\aDark = \aDarkLow$.
Depending on the \PZprime mass, $0.07 < \rinv < 0.53$ ($0.06 < \rinv < 0.57$) and all \mDark and \aDark variations considered are also observed (expected) to be excluded.
These exclusions do not rely strongly on the details of the jet substructure, so they can apply to many signal models that satisfy the final selection via a resonance that produces jets aligned with missing transverse momentum.
In particular, because the BDT efficiency depends on \mDark, the inclusive search provides stronger limits at very high and very low \mDark values.
Within the excluded range of \rinv values, the corresponding mass exclusion is more stringent than the DM mediator interpretation from the inclusive dijet search~\cite{Sirunyan:2019vgj}, which targets the \PZprime decay to SM quarks.
In the HV model considered here, \Bdark is similar to the branching fraction for ${\PZprime \to \PQq\PAQq}$,
leading the two strategies to have similar sensitivities in regions where they have similar acceptance;
for larger \Bdark values, the relative advantage of the dedicated search strategy would increase.
For signal models with very low or very high \rinv values, outside of the excluded range, the efficiency of the selection in this search decreases.
The dijet resonance search~\cite{Sirunyan:2019vgj} and searches based on missing transverse momentum recoiling against an ISR jet~\cite{CMS:2021far} provide better sensitivity for such signal models.

The results from the BDT-based signal regions increase the observed (expected) excluded mediator mass range to $1.5 < \mZprime < 5.1\TeV$ ($1.5 < \mZprime < 5.1\TeV$) for wide ranges of the other signal parameters.
The range of observed (expected) excluded \rinv values also increases to $0.01 < \rinv < 0.77$ ($0.01 < \rinv < 0.78$),
and the \mDark and \aDark variations are excluded for a wider range of \PZprime masses.
The observed upper limit on the mediator mass decreases compared to the expected limit in some regions of the signal model parameter space
because of a small excess with a significance of 1--2 standard deviations around $\mZprime = 3.9\TeV$.
The improved exclusions from the BDT apply strictly to signal models that produce dark showers with the characteristics described in Sections~\ref{sec:samples}~and~\ref{sec:tagger}.

\section{Summary}\label{sec:summary}

We present the first collider search for resonant production of dark matter from a strongly coupled hidden sector.
The search uses proton-proton collision data collected with the CMS detector in 2016--2018,
corresponding to an integrated luminosity of 138\fbinv at a center-of-mass energy of 13\TeV.
The signal model introduces a dark sector with multiple flavors of dark quarks that are charged under a dark confining force and form stable and unstable dark hadrons.
The stable dark hadrons constitute dark matter candidates, while the unstable dark hadrons decay promptly to standard model (SM) quarks,
forming collimated sprays of both invisible and visible particles known as ``semivisible'' jets.
The hidden sector communicates with the SM via a leptophobic \PZprime boson mediator.
We consider variations in several parameters of the hidden sector signal model:
the \PZprime mass, \mZprime; the dark hadron mass, \mDark; the fraction of stable, invisible dark hadrons, \rinv; and the running coupling of the dark force, \aDark.

We pursue a dual strategy to maximize both generality and sensitivity.
The general version of the search, which uses only event-level kinematic variables,
excludes models with $1.5 < \mZprime < 4.0\TeV$ and $0.07 < \rinv < 0.53$ at 95\% confidence level (\CL), depending on the other signal model parameters.
The more sensitive version of the search employs a boosted decision tree (BDT) to discriminate between signal and background jets.
With the BDT-based search, the 95\% \CL exclusion limits extend to $1.5 < \mZprime < 5.1\TeV$ and $0.01 < \rinv < 0.77$, for wider ranges of the other signal model parameters.
Depending on the mediator mass, all variations considered for \mDark and \aDark can be excluded.
These improvements indicate the power of machine learning techniques to separate dark sector signals from large SM backgrounds.

These results complement existing searches for dijet resonances and dark matter events with missing momentum and initial-state radiation.
Existing strategies did not target strongly coupled hidden sector models, which produce events with jets aligned with moderate missing transverse momentum.
Compared to standard dijet searches, the backgrounds are reduced and the resolution is improved by including missing momentum in the event selection and the resonance mass reconstruction, respectively.
Events with jets aligned with missing momentum are explicitly rejected from other collider dark matter searches.
In addition, the use of jet substructure provides a substantial increase in model-dependent sensitivity.
As a result, a wide range of these models with intermediate \rinv values are now excluded for the first time.

\begin{acknowledgments}
        We congratulate our colleagues in the CERN accelerator departments for the excellent performance of the LHC and thank the technical and administrative staffs at CERN and at other CMS institutes for their contributions to the success of the CMS effort. In addition, we gratefully acknowledge the computing centers and personnel of the Worldwide LHC Computing Grid and other centers for delivering so effectively the computing infrastructure essential to our analyses. Finally, we acknowledge the enduring support for the construction and operation of the LHC, the CMS detector, and the supporting computing infrastructure provided by the following funding agencies: BMBWF and FWF (Austria); FNRS and FWO (Belgium); CNPq, CAPES, FAPERJ, FAPERGS, and FAPESP (Brazil); MES and BNSF (Bulgaria); CERN; CAS, MoST, and NSFC (China); MINCIENCIAS (Colombia); MSES and CSF (Croatia); RIF (Cyprus); SENESCYT (Ecuador); MoER, ERC PUT and ERDF (Estonia); Academy of Finland, MEC, and HIP (Finland); CEA and CNRS/IN2P3 (France); BMBF, DFG, and HGF (Germany); GSRI (Greece); NKFIA (Hungary); DAE and DST (India); IPM (Iran); SFI (Ireland); INFN (Italy); MSIP and NRF (Republic of Korea); MES (Latvia); LAS (Lithuania); MOE and UM (Malaysia); BUAP, CINVESTAV, CONACYT, LNS, SEP, and UASLP-FAI (Mexico); MOS (Montenegro); MBIE (New Zealand); PAEC (Pakistan); MSHE and NSC (Poland); FCT (Portugal); JINR (Dubna); MON, RosAtom, RAS, RFBR, and NRC KI (Russia); MESTD (Serbia); MCIN/AEI and PCTI (Spain); MOSTR (Sri Lanka); Swiss Funding Agencies (Switzerland); MST (Taipei); ThEPCenter, IPST, STAR, and NSTDA (Thailand); TUBITAK and TAEK (Turkey); NASU (Ukraine); STFC (United Kingdom); DOE and NSF (USA).
        
        \hyphenation{Rachada-pisek} Individuals have received support from the Marie-Curie program and the European Research Council and Horizon 2020 Grant, contract Nos.\ 675440, 724704, 752730, 758316, 765710, 824093, 884104, and COST Action CA16108 (European Union); the Leventis Foundation; the Alfred P.\ Sloan Foundation; the Alexander von Humboldt Foundation; the Belgian Federal Science Policy Office; the Fonds pour la Formation \`a la Recherche dans l'Industrie et dans l'Agriculture (FRIA-Belgium); the Agentschap voor Innovatie door Wetenschap en Technologie (IWT-Belgium); the F.R.S.-FNRS and FWO (Belgium) under the ``Excellence of Science -- EOS" -- be.h project n.\ 30820817; the Beijing Municipal Science \& Technology Commission, No. Z191100007219010; the Ministry of Education, Youth and Sports (MEYS) of the Czech Republic; the Deutsche Forschungsgemeinschaft (DFG), under Germany's Excellence Strategy -- EXC 2121 ``Quantum Universe" -- 390833306, and under project number 400140256 - GRK2497; the Lend\"ulet (``Momentum") Program and the J\'anos Bolyai Research Scholarship of the Hungarian Academy of Sciences, the New National Excellence Program \'UNKP, the NKFIA research grants 123842, 123959, 124845, 124850, 125105, 128713, 128786, and 129058 (Hungary); the Council of Science and Industrial Research, India; the Latvian Council of Science; the Ministry of Science and Higher Education and the National Science Center, contracts Opus 2014/15/B/ST2/03998 and 2015/19/B/ST2/02861 (Poland); the Funda\c{c}\~ao para a Ci\^encia e a Tecnologia, grant CEECIND/01334/2018 (Portugal); the National Priorities Research Program by Qatar National Research Fund; the Ministry of Science and Higher Education, projects no. 14.W03.31.0026 and no. FSWW-2020-0008, and the Russian Foundation for Basic Research, project No.19-42-703014 (Russia); MCIN/AEI/10.13039/501100011033, ERDF ``a way of making Europe", and the Programa Estatal de Fomento de la Investigaci{\'o}n Cient{\'i}fica y T{\'e}cnica de Excelencia Mar\'{\i}a de Maeztu, grant MDM-2017-0765 and Programa Severo Ochoa del Principado de Asturias (Spain); the Stavros Niarchos Foundation (Greece); the Rachadapisek Sompot Fund for Postdoctoral Fellowship, Chulalongkorn University and the Chulalongkorn Academic into Its 2nd Century Project Advancement Project (Thailand); the Kavli Foundation; the Nvidia Corporation; the SuperMicro Corporation; the Welch Foundation, contract C-1845; and the Weston Havens Foundation (USA).
\end{acknowledgments}

\bibliography{auto_generated}
\cleardoublepage \appendix\section{The CMS Collaboration \label{app:collab}}\begin{sloppypar}\hyphenpenalty=5000\widowpenalty=500\clubpenalty=5000\cmsinstitute{Yerevan~Physics~Institute, Yerevan, Armenia}
A.~Tumasyan
\cmsinstitute{Institut~f\"{u}r~Hochenergiephysik, Vienna, Austria}
W.~Adam\cmsorcid{0000-0001-9099-4341}, J.W.~Andrejkovic, T.~Bergauer\cmsorcid{0000-0002-5786-0293}, S.~Chatterjee\cmsorcid{0000-0003-2660-0349}, K.~Damanakis, M.~Dragicevic\cmsorcid{0000-0003-1967-6783}, A.~Escalante~Del~Valle\cmsorcid{0000-0002-9702-6359}, R.~Fr\"{u}hwirth\cmsAuthorMark{1}, M.~Jeitler\cmsAuthorMark{1}\cmsorcid{0000-0002-5141-9560}, N.~Krammer, L.~Lechner\cmsorcid{0000-0002-3065-1141}, D.~Liko, I.~Mikulec, P.~Paulitsch, F.M.~Pitters, J.~Schieck\cmsAuthorMark{1}\cmsorcid{0000-0002-1058-8093}, R.~Sch\"{o}fbeck\cmsorcid{0000-0002-2332-8784}, D.~Schwarz, S.~Templ\cmsorcid{0000-0003-3137-5692}, W.~Waltenberger\cmsorcid{0000-0002-6215-7228}, C.-E.~Wulz\cmsAuthorMark{1}\cmsorcid{0000-0001-9226-5812}
% ...
\vskip\cmsinstskip
% ...
1:~Also at TU Wien, Wien, Austria\\
% ...
\end{sloppypar}
\end{document}